\documentclass[twocolumn,twocolappendix,resetfootnote]{aastex701}  

\usepackage{amsmath}
\usepackage{graphicx}
\usepackage{threeparttable}
\usepackage{txfonts}
\let\oldtextsc\textsc
\renewcommand{\textsc}[1]{\textup{\oldtextsc{#1}}}
\usepackage{xcolor}
   
\hypersetup{allcolors=blue}

\graphicspath{{./Figures/}{./figures/}}

\newcommand{\kms}{\,\ensuremath{\mathrm{km\,s^{-1}}}}
\newcommand{\msun}{{$M_{\odot}$}}

\newcommand{\gtsima}{$\; \buildrel > \over \sim \;$}
\newcommand{\ltsima}{$\; \buildrel < \over \sim \;$}
\newcommand{\prosima}{$\; \buildrel \propto \over \sim \;$}
\newcommand{\gsim}{\lower.5ex\hbox{\gtsima}}
\newcommand{\lsim}{\lower.5ex\hbox{\ltsima}}
\newcommand{\simgt}{\lower.5ex\hbox{\gtsima}}
\newcommand{\simlt}{\lower.5ex\hbox{\ltsima}}
\newcommand{\simpr}{\lower.5ex\hbox{\prosima}}

\newcommand\iona[2]{#1$\;${\scshape{#2}}}

\renewcommand{\footnotesize}{\small}

\begin{document}

\title{DAJE-JWST Spectroscopy of the Lensed Little Red Dot A2744-QSO1: A Massive Black Hole Enshrouded in Dense Gas}

\author[0000-0002-4436-6923]{Fan Zou}\email{fanzou01@gmail.com}
\affiliation{Department of Astronomy, University of Michigan, 1085 S. University Ave, Ann Arbor, MI 48109, USA}
\author[0000-0001-5802-6041]{Elena Gallo}\email{egallo@umich.edu}
\affiliation{Department of Astronomy, University of Michigan, 1085 S. University Ave, Ann Arbor, MI 48109, USA}
\author[0000-0002-8460-0390]{Tommaso Treu}\email{tt@astro.ucla.edu}
\affiliation{Department of Physics and Astronomy, University of California Los Angeles, 430 Portola Plaza, Los Angeles, CA 90095-1547, USA}
\author[0000-0003-3820-2823]{Adriano Fontana}\email{adriano.fontana@inaf.it}
\affiliation{INAF - Osservatorio Astronomico di Roma, via di Frascati 33, 00078 Monte Porzio Catone, Italy}
\author[0000-0002-4140-1367]{Guido Roberts-Borsani}\email{g.robertsborsani@ucl.ac.uk}
\affiliation{Department of Physics \& Astronomy, University College London, Gower Street, London WC1E 6BT, UK}
\author[0009-0000-8489-7211]{Raunaq Singh Rai}\email{raunaq.rai.25@ucl.ac.uk}
\affiliation{Department of Physics \& Astronomy, University College London, Gower Street, London WC1E 6BT, UK}
\author[0000-0002-9935-6047]{Massimo Stiavelli}\email{mstiavel@stsci.edu}
\affiliation{Space Telescope Science Institute, 3700 San Martin Dr., 21218 Baltimore, USA}
\author[0009-0001-9416-0923]{Giorgia Di Rosa}\email{giorgia.dirosa@unife.it}
\affiliation{Dipartimento di Fisica e Scienze della Terra, Universit\`a degli Studi di Ferrara, Via G.Saragat 1, 44122 Ferrara, Italy}
\author[0000-0002-5116-7287]{Daniel Gilman}\email{gilmanda@uchicago.edu}
\affiliation{Kavli Institute for Cosmological Physics, University of Chicago, Chicago, IL 60637, USA}
\author[0000-0002-3336-4965]{Marco Lombardi}\email{marco.lombardi@unimi.it}
\affiliation{Dipartimento di Fisica ``Aldo Pontremoli'', Universit\`a degli Studi di Milano, Via Celoria 16, I-20133 Milano, Italy}
\author[0000-0002-6813-0632]{Piero Rosati}\email{rstpri1@unife.it}
\affiliation{Dipartimento di Fisica e Scienze della Terra, Universit\`a degli Studi di Ferrara, Via G.Saragat 1, 44122 Ferrara, Italy}
\author[0000-0002-6909-205X]{Paras Sharma}\email{paras.sharma@stonybrook.edu}
\affiliation{Department of Physics and Astronomy, Stony Brook University, Stony Brook, NY 11794, USA}
\author[0000-0003-3108-9039]{Ana Acebron}\email{ana.acebron@unican.es}
\affiliation{Instituto de F\'isica de Cantabria (CSIC-UC). Avda. Los Castros s/n. 39005 Santander, Spain}
\author[0000-0003-0727-0137]{Lorenzo Bazzanini}\email{lorenzo.bazzanini@inaf.it}
\affiliation{INAF - IRA, Istituto di Radioastronomia, via Gobetti 101, I-40129, Bologna, Italy}
\author[0000-0003-1383-9414]{Pietro Bergamini}\email{pietro.bergamini@inaf.it}
\affiliation{INAF - OAS, Osservatorio di Astrofisica e Scienza dello Spazio di Bologna, via Gobetti 93/3, I-40129 Bologna, Italy}
\author[0000-0001-5984-0395]{Maru\v{s}a Brada\v{c}}\email{marusa.bradac@fmf.uni-lj.si}
\affiliation{University of Ljubljana, Faculty of Mathematics and Physics, Jadranska ulica 19, SI-1000 Ljubljana, Slovenia}
\affiliation{Department of Physics and Astronomy, University of California Davis, 1 Shields Avenue, Davis, CA 95616, USA}
\author[0000-0003-2536-1614]{Antonello Calabr\`{o}}\email{antonello.calabro@inaf.it}
\affiliation{INAF - Osservatorio Astronomico di Roma, via di Frascati 33, 00078 Monte Porzio Catone, Italy}
\author[0009-0008-0116-4716]{Riccardo Caleno}\email{riccardo.caleno@uniroma1.it}
\affiliation{Dipartimento di Fisica, Sapienza Universit\`a di Roma, Piazzale Aldo Moro 2, 00185 Roma, Italy}
\author[0000-0001-9875-8263]{Marco Castellano}\email{marco.castellano@inaf.it}
\affiliation{INAF - Osservatorio Astronomico di Roma, via di Frascati 33, 00078 Monte Porzio Catone, Italy}
\author[0000-0002-2912-3923]{Sunmyon Chon}\email{sunmyon.chon@roma1.infn.it}
\affiliation{Dipartimento di Fisica, Sapienza Universit\`a di Roma, Piazzale Aldo Moro 2, 00185 Roma, Italy}
\affiliation{Astronomical Institute, Graduate School of Science, Tohoku University, 6--3 Aoba, Sendai 980-8578, Japan}
\author[0000-0001-7782-7071]{Richard Ellis}\email{richard.ellis@ucl.ac.uk}
\affiliation{Department of Physics \& Astronomy, University College London, Gower Street, London WC1E 6BT, UK}
\author[0000-0003-3248-5666]{Giovanni Gandolfi}\email{giovigandolfi@gmail.com}
\affiliation{INAF - Osservatorio Astronomico di Roma, via di Frascati 33, 00078 Monte Porzio Catone, Italy}
\author[0000-0002-3254-9044]{Karl Glazebrook}\email{kglazebrook@swin.edu.au}
\affiliation{JWST Australian Data Centre, Centre for Astrophysics and Supercomputing, Swinburne University of Technology, PO Box 218, Hawthorn, VIC 3122, Australia}
\author[0000-0002-5926-7143]{Claudio Grillo}\email{claudio.grillo@unimi.it}
\affiliation{Dipartimento di Fisica ``Aldo Pontremoli'', Universit\`a degli Studi di Milano, Via Celoria 16, I-20133 Milano, Italy}
\affiliation{INAF -- IASF Milano, via A. Corti 12, I-20133 Milano, Italy}
\author[0000-0001-6919-1237]{Matthew Malkan}\email{malkan@astro.ucla.edu}
\affiliation{Department of Physics and Astronomy, University of California Los Angeles, 430 Portola Plaza, Los Angeles, CA 90095-1547, USA}
\author[0000-0003-1225-7084]{Massimo Meneghetti}\email{massimo.meneghetti@inaf.it}
\affiliation{INAF - OAS, Osservatorio di Astrofisica e Scienza dello Spazio di Bologna, via Gobetti 93/3, I-40129 Bologna, Italy}
\author[0000-0003-1427-2456]{Matteo Messa}\email{matteo.messa@inaf.it}
\affiliation{INAF - OAS, Osservatorio di Astrofisica e Scienza dello Spazio di Bologna, via Gobetti 93/3, I-40129 Bologna, Italy}
\author[0000-0002-8512-1404]{Takahiro Morishita}\email{morishita@astr.tohoku.ac.jp}
\affiliation{Astronomical Institute, Graduate School of Science, Tohoku University, 6--3 Aoba, Sendai 980-8578, Japan}
\author[0000-0003-2804-0648]{Themiya Nanayakkara}\email{themiya.nanayakkara@sydney.edu.au}
\affiliation{Sydney Institute for Astronomy, School of Physics A28, The University of Sydney, NSW 2006, Australia.}
\author[0000-0001-5595-757X]{Evan Nunez-Cravin}\email{evanhazey@gmail.com}
\affiliation{The Observatories of the Carnegie Institution for Science, 813 Santa Barbara Street, Pasadena, CA, USA}
\author[0000-0002-0939-9156]{Borja P\'{e}rez-D\'{i}az}\email{borja.perezdiaz@inaf.it}
\affiliation{INAF - Osservatorio Astronomico di Roma, via di Frascati 33, 00078 Monte Porzio Catone, Italy}
\author[0000-0002-9334-8705]{Paola Santini}\email{paola.santini@inaf.it}
\affiliation{INAF - Osservatorio Astronomico di Roma, via di Frascati 33, 00078 Monte Porzio Catone, Italy}
\author[0000-0001-9317-2888]{Raffaella Schneider}\email{raffaella.schneider@uniroma1.it}
\affiliation{Dipartimento di Fisica, Sapienza Universit\`a di Roma, Piazzale Aldo Moro 2, 00185 Roma, Italy}
\author[0000-0001-9391-305X]{Michele Trenti}\email{michele.trenti@unimelb.edu.au}
\affiliation{School of Physics, The University of Melbourne, VIC 3010, Australia}
\author[0000-0002-5057-135X]{Eros Vanzella}\email{eros.vanzella@inaf.it}
\affiliation{INAF - OAS, Osservatorio di Astrofisica e Scienza dello Spazio di Bologna, via Gobetti 93/3, I-40129 Bologna, Italy}
\author[0000-0003-0980-1499]{Benedetta Vulcani}\email{benedetta.vulcani@inaf.it}
\affiliation{INAF - Osservatorio Astronomico di Padova, Vicolo Osservatorio 5, 35122 Padova, Italy}
\author[0000-0003-4244-8527]{Tommaso Zana}\email{tommaso.zana@uniroma1.it}
\affiliation{Dipartimento di Fisica, Sapienza Universit\`a di Roma, Piazzale Aldo Moro 2, 00185 Roma, Italy}
\affiliation{INAF - Osservatorio Astronomico di Roma, via di Frascati 33, 00078 Monte Porzio Catone, Italy}

\begin{abstract}
We investigate A2744-QSO1, a triply lensed little red dot
at $z=7.04$, with new ultra-deep NIRSpec spectroscopy at $R\sim1000$
from the Deep Amplified JWST Experiment (DAJE) and archival spectra.
Accounting for the lensing time delays, microlensing and achromatic
millilensing, we find no significant variability of the H$\beta$ and
H$\alpha$ equivalent widths in seven spectra spanning 2.8 rest-frame
years. The variability reported earlier lies within the
systematic uncertainties of the measurements, while our limits (15\% in
H$\alpha$) still allow the $\lesssim7\%$ changes expected for an
ordinary quasar. The Balmer line profiles reveal an intermediate-width
H$\alpha$ component (${\rm FWHM}\simeq700$\,km\,s$^{-1}$),  
with no detected counterpart in H$\beta$, which likely
arises outside the broad-line region.
Once this component is excluded, broad H$\beta$ and H$\alpha$ have
comparable widths (${\rm FWHM}\simeq2000$\,km\,s$^{-1}$) and imply a
black hole mass of $10^{6.7}$--$10^{7.3}\,$M$_\odot$. Within the
0.4\,dex scatter of single-epoch estimators, this mass agrees with the
dynamical estimate of $10^{7.7\pm0.3}\,$M$_\odot$, as expected if the
lines are broadened by gravity rather than by electron scattering. We
detect narrow He\,\textsc{i}~$\lambda5876$; its ratio to narrow
H$\beta$ is several times the recombination value, requiring very
dense gas ($n_e\gtrsim10^{9}$\,cm$^{-3}$) with a small velocity
dispersion, outside the broad-line region. The broad Balmer decrement
implies either 1.5--3 magnitudes of visual extinction or gas denser
than $10^{8.5}$\,cm$^{-3}$. Photoionization models of the narrow lines
yield $12+\log({\rm O/H})=6.0$--7.1 (0.2--3\% of the solar value), at
which little dust is expected to form. These results point to a
massive black hole enshrouded in dense, metal-poor gas.
\end{abstract}


\section{Introduction}
\label{sec:intro}

Among the most unexpected findings of the James Webb Space Telescope
(JWST) is a population of compact (half-light radius
$r_{\rm e}\lesssim100$\,pc), very red sources at $z\gtrsim4$, commonly
referred to as ``little red dots'' (LRDs; \citealt{matthee2024,
kocevski2025, greene2024, labbe2025, kokorev2024}), whose rest-frame
UV-to-optical spectral energy distribution (SED) is V-shaped, with the
inflection near the Balmer limit. Rest-frame optical spectroscopy
reveals broad Balmer lines in most of them, with widths of
$\sim10^{3}$\,\kms; if these trace virial motion, they imply black
hole masses of $M_{\rm BH}\sim10^{6}$--$10^{8}$\,\msun\
\citep{matthee2024, kocevski2025}. Many LRDs also show narrow Balmer
absorption and a strong Balmer break, which suggest that the black
hole is enshrouded in dense gas \citep{wang2024, inayoshi2025b,
naidu2025, degraaff2025}, possibly growing rapidly through
super-Eddington accretion.

The nature of LRDs, however, remains unsettled. If they are powered by accretion, their properties appear at odds with those of dust-reddened active galactic nuclei (AGN):
they lack X-ray emission,
both individually and in deep stacks \citep{yue2024, ananna2024,
maiolino2025}, and radio emission \citep{mazzolari2026radio,
perger2025}; they show no hot dust emission in the mid-infrared
\citep{setton2025}; their continuum and broad lines show weak or no
variability on short rest-frame timescales \citep{kokubo2025,
stone2026, brooks2026}; and their He\,\textsc{ii} emission is weak
\citep{wang2026a, wang2026b}.

To explain these observations, two broad classes of models have emerged. In the first, the accreting black hole is enshrouded in dense (hydrogen density $n_{\rm H}\sim10^{9}$--$10^{11}$\,cm$^{-3}$), nearly dust-free gas, which produces the observed Balmer break and absorption, as well as the red optical continuum, through collisional excitation and photospheric-like emission at an effective temperature $T_{\rm eff}\approx5000$\,K \citep{inayoshi2025b, naidu2025, degraaff2025, ji2025, rusakov2026}. In the second, the black hole accretes at super-Eddington rates through a geometrically thick gas flow or envelope, which in turn suppresses the X-ray emission \citep{pacucci2024, kido2025, madau2025}. In the shadowing variant of this model, the thick inner disk shields 
the broad-line region from the hardest ionizing photons, and dusty 
clouds near the equatorial plane redden the broad lines 
\citep{madau2026a}. Other proposals place a direct-collapse black hole inside a dense inflow of nearly pristine gas \citep{chon2026a, pacucci2026}, or a super-Eddington engine of intermediate mass inside a dense wind \citep{naidu2026}. A joint analysis of the broad Balmer lines and X-ray limits of JWST AGNs favors low-mass black holes accreting above the Eddington rate \citep{Trinca2026}, and hydrodynamical simulations show that such episodes can be sustained for tens of Myr, with their duration and the resulting growth set by the gas supply and black hole feedback \citep{Lupi2024, Zana2026, Caleno2026}.

The models can be distinguished by three observables: the ionizing spectrum that reaches the line-emitting gas, the origin of emission line widths, and the timescales of black hole variability. An accretion disk radiating below the Eddington limit varies in the same fashion as quasars: Sloan Digital Sky Survey (SDSS) quasars vary by $\simeq0.2$\,mag in the rest-frame ultraviolet over baselines of a few hundred days or more \citep{macleod2010}. In the envelope and wind models, by contrast, photons diffuse through the dense gas before escaping, which predicts
variability on timescales of years or longer  \citep{secunda2025, sneppen2026, naidu2026}. Continuum, broad and narrow lines carry complementary information. 
The continuum traces the accretion flow directly; the broad lines 
follow it with a lag set by the light-crossing time of the broad-line 
region (light-days for $M_{\rm BH}\sim10^{7}$\,\msun) and with a 
smaller amplitude, so their EWs change by only a fraction of the 
continuum variation; the narrow lines arise over $\gtrsim100$\,pc, 
are constant on these timescales, and provide an internal flux 
reference.\\

A2744-QSO1 (hereafter QSO1) is arguably the most extensively studied LRD to date. It was discovered in JWST Near Infrared Camera (NIRCam) imaging of the Abell~2744 cluster field, by the UNCOVER program \citep{furtak2023}, as an unresolved ($r_{\rm e}<30$\,pc), extremely red and UV-faint source, triply imaged by the foreground cluster. One of its images is also detected in GLASS-JWST Near Infrared Imager and Slitless Spectrograph (NIRISS) direct imaging \citep{watson2025}. Spectroscopy of all three images with the Near Infrared Spectrograph (NIRSpec) revealed broad Balmer lines, implying a black hole of a few $10^{7}$\,\msun\ \citep{furtak2024}, and the source is undetected in deep Chandra imaging \citep{wangs2026clues}. Higher-resolution spectroscopy resolved a non-stellar Balmer break and narrow Balmer absorption \citep{ji2025, deugenio2025}. Integral Field Unit (IFU) observations revealed Keplerian rotation in the broad-line region, yielding a dynamical mass of
$10^{7.7\pm0.3}$\,\msun, consistent with the single epoch virial estimate and about
two orders of magnitude above that implied by electron-scattering
broadening \citep{juodzbalis2026, deugenio2025, rusakov2026}. The Eddington ratio inferred for QSO1 ranges from a few percent to a few tens of percent depending on the mass adopted \citep{furtak2024, deugenio2025, juodzbalis2026}, well below the rates assumed in super-Eddington models \citep{madau2026a}. In the rest-frame ultraviolet, QSO1 shows broad Ly$\alpha$ and no broad He\,\textsc{ii}~$\lambda$1640 \citep{tang2026}.

\begin{figure}
\centering
\includegraphics[width=\columnwidth]{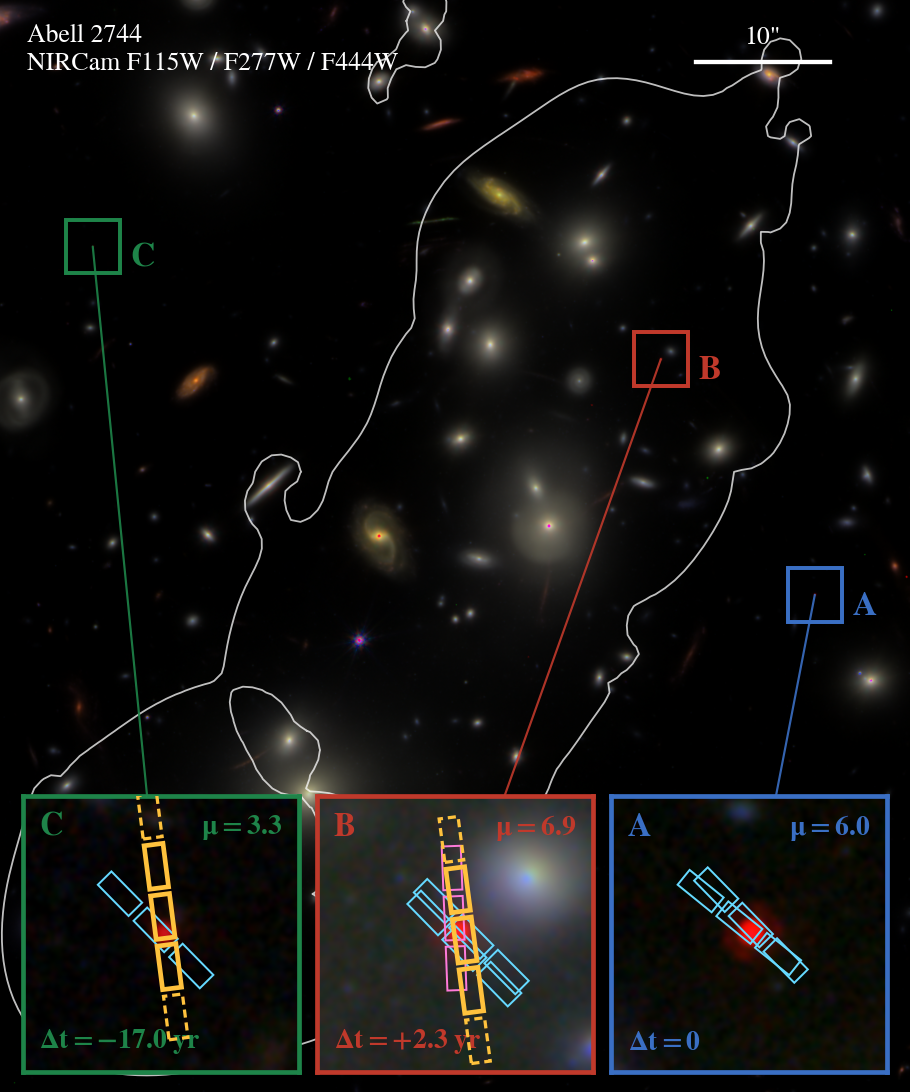}
\caption{The three lensed images of QSO1 in Abell~2744. JWST/NIRCam
F115W/F277W/F444W composite of a $68''\times82''$ field (UNCOVER and
MegaScience mosaics, v7.2; \citealt{bezanson2024,suess2024}); north
up and east left. The white curve is the critical line of model~v1.9
(G. Di Rosa et al., in prep.) at the redshift of QSO1. The $2\farcs9$
cutouts show the three images, labeled with the observed-frame
arrival delay $\Delta t$ of each image relative to A and with its
magnification (Section~\ref{sec:epochs}), and the open areas of the
MSA shutters of the spectra used in this paper: DAJE at the first nod
position (thick solid yellow) and the additional shutters used at
nods 2 and 3 (dashed yellow); UNCOVER (thin blue; in 2023 two
configurations covering A and B and one covering C, in 2024 one
covering A); and SPURS (thin pink). The archival shutters are shown
at the first dither position; the gaps between shutters are the bars.
The DAJE spectrum of image A is contaminated and is not used in this
work (Section~\ref{sec:data}), so its shutters are not shown. The
DAJE spectra of images B and C are shown in Figure~\ref{fig:lines}.}
\label{fig:showpiece}
\end{figure}

QSO1 is one of four known multiply imaged LRDs \citep{zhang2025} and
the only one with spectroscopy of every image. The lensing delays span
2.4\,yr in the rest frame, so a single observation samples the source
at three epochs (Figure~\ref{fig:showpiece}). From the UNCOVER PRISM
spectra, \citet{furtak2025} and \citet{ji2025} report lower
Balmer-line equivalent widths (EWs) in image A than in image C, while
the continuum flux ratio of the two images is the same from the near-
to the mid-infrared \citep{iani2026}. If intrinsic, the difference would indicate variability of the broad 
lines over about two rest-frame years. The three images of QSO1 thus 
offer a direct test of the variability predicted by the models. 

Here we present ultra-deep NIRSpec Micro-Shutter Assembly (MSA)
spectroscopy of the three images of QSO1, obtained by the Deep
Amplified JWST Experiment (DAJE-JWST) program\footnote{\url{https://www.daje-jwst.com}}
(JWST Cycle~5 GO~9645; PIs T.~Treu, A.~Fontana, G.~Roberts-Borsani;
T.~Treu et al., 2026, in prep.), with total exposure times of up to
134\,h per image summed over all gratings. We analyze images B and C
(Section~\ref{sec:data}).

We combine the new data with archival spectroscopy to carry out four 
empirical tests, each aimed at one of the distinctions between the 
models above. First, we test whether the reported variability is 
intrinsic by comparing the EWs of H$\beta$ and H$\alpha$ in seven 
spectra that sample the source over 2.8 rest-frame years, including 
images B and C, 2.4 rest-frame yr apart in the DAJE data; a disk 
below the Eddington limit should vary as quasars do, whereas photon 
diffusion through an envelope smooths the variations over years. 
Second, we compare the profiles of broad H$\beta$ and H$\alpha$ to 
establish which parts of the lines arise in the broad-line region and 
what black hole mass they imply; if the lines are broadened by 
gravity, this mass should agree with the dynamical one, while 
electron scattering would imply a mass about two orders of magnitude 
lower. Third, we use the Balmer decrement, the higher-order Balmer 
lines and He\,\textsc{i}~$\lambda5876$ to test whether the 
line-emitting gas is dense or reddened by dust, as in the dense-gas 
and shadowing models, respectively. Fourth, we measure the oxygen abundance of the narrow-line gas, to 
test whether the black hole is embedded in nearly pristine gas, as in 
direct-collapse scenarios.

This paper is organized as follows. Section~\ref{sec:data} describes the observations and data reduction, Section~\ref{sec:lensing} the lens model, the flux-ratio anomaly among images, and its interpretation in terms of microlensing and millilensing. Section~\ref{sec:analysis} discusses the spectral analysis, while Section~\ref{sec:discussion} discusses the origin of the line widths, the physical state of the line-emitting gas and the metallicity of its surroundings. We summarize our results and present our conclusions in Section~\ref{sec:summary}. Throughout the paper we adopt a flat $\Lambda$CDM cosmology with  $H_0=70$\,km\,s$^{-1}$\,Mpc$^{-1}$ and $\Omega_{\rm m}=0.3$, and the systemic redshift $z_{\rm sys}=7.0364$ measured from the narrow [O\,\textsc{iii}] lines by \citet{tang2026}. 

%%%%%%%% Figure 1 %%%%%%%%%%%%%%%%%%%%%%%%%%%%%%%%%%%%%%%%%%%
%%%%%%%%%%%%%%%%%%%%%%%%%
\section{Observations and data reduction}
\label{sec:data}
The three images of QSO1 were observed between 2026 November 16--24 (MJD 61361--61368) in a single MSA
configuration of the DAJE program, at a resolving power
$R\equiv\lambda/\Delta\lambda\simeq1000$, using the G140M, G235M and G395M gratings with the
F100LP, F170LP and F290LP blocking filters, which together cover
$\simeq$0.97--5.5\,$\mu$m. Images B and C were observed through three-shutter slitlets at all three nod
positions, for a total on-source time of 50.8, 45.2 and 38.2\,h in G140M, G235M and G395M (1.7, 5.6 and 13 times the exposures of the deepest previous medium-resolution spectrum of QSO1; \citealt{tang2026}). Image A instead fell
in a two-shutter slitlet due to stuck-open and stuck-closed shutters, for a total on-source time of 33.9, 30.1 and 25.5\,h, respectively. Since the three images are observed simultaneously in one MSA
configuration, the data sample QSO1 at epochs separated by the
lensing delays. This removes the changes in calibration
between visits that limit multi-epoch MSA studies \citep{stone2026,
brooks2026}. The images still differ, however, in magnification, slit losses, contamination and
microlensing; Sections~\ref{sec:lensing} and \ref{sec:fitting} describe how each is accounted for. 
The delay between images A and B is $104^{+4}_{-3}$\,d in the rest frame, and that between C and A
$772^{+16}_{-18}$\,d (G. Di Rosa et al., in prep.; also see Section~\ref{sec:epochs} for the lensing model). Images B and C, the two analyzed here, are 876\,d apart, longer than the light-crossing time of the broad-line 
region of QSO1 \citep{bentz2013}, and thus probe the broad lines after they have had
time to respond to a change in the continuum. The archival spectra utilized here are the UNCOVER PRISM spectra (GO~2561; \citealt{bezanson2024}) of the three images, with a second epoch of image A \citep{furtak2024, furtak2025}, and the SPURS spectra (GO~9214; PIs Mason \& Stark) of image B \citep{tang2026}. To extend the
comparison to earlier epochs with as few systematics as possible,
we re-reduce them with the same pipeline, and fit them in the same way as the DAJE data. The DAJE spectrum of image A could not be used due to significant spectral contamination from a nearby source. 
Since the MSA configuration is the same in all visits, we defer the analysis of image A to a future paper with a forward model of the contamination, and analyze the DAJE spectra of images B and C.

All the data were reduced and the spectra extracted as described in
\citet{rai2026a}, using the \texttt{msaexp} code \citep{brammer2023} (version 0.9.18) which adopts the standard STScI pipeline \citep{Bushouse2025} (version 3.0.0), with CRDS context jwst\_1623.pmap. With the extended wavelength range compared to previous version of the pipeline, G395M reaches $\approx5.5\,\mu$m and thus includes H$\alpha$ at 5.276\,$\mu$m. We refer the reader to the aforementioned paper for details, however in brief: After correcting for detector-level artifacts, each 2D exposure is background-subtracted with the nodded exposures in the adjacent shutters; the one-dimensional spectrum is then optimally extracted \citep{horne1986} with a Gaussian kernel fitted to the spatial profile of the source spectrum, and the resulting exposures are then combined with inverse-variance weights. The resulting spectra of images B and C are shown in
Figure~\ref{fig:lines}. 

Due to an increasing point spread function (PSF) with wavelength, some total and wavelength-dependent flux losses are expected due to the small size of the MSA shutters. The flux losses are corrected following \citet{rai2026b}, by forward-modeling the light of a morphological model through the MSA shutter with the \texttt{msafit} \citep{degraaff2024msafit} code. The model is constructed with \texttt{PySersic} \citep{pasha2023}, by jointly fitting to all available NIRCam imaging (adopting the UNCOVER mosaics, free from intra-cluster and cluster galaxy light), and a wavelength-dependent throughput is derived by comparing the total-to-shutter in each band. A wavelength-dependent flux loss correction relative to a centered point source is then determined with a fourth-order polynomial fit. Table~\ref{tab:epochs} lists the correction at H$\beta$ and H$\alpha$
for each spectrum; their ratio enters H$\alpha$/H$\beta$, while the EWs
are independent of it.
Different slit-loss models (e.g., \texttt{msaexp}, \texttt{msafit} and a point-source model) differ by up to 15\% in H$\alpha$/H$\beta$ and by up to 25\% in the flux of image C relative to B; we include the first as a systematic uncertainty on the Balmer decrement (Section~\ref{sec:decrement}).
Since the images are unresolved \citep{furtak2023}, the slit losses
cancel in the EWs, provided the line and the continuum have the same
spatial distribution on the scale of a shutter. This holds for the
continuum and the broad lines; the narrow lines, which are spatially
resolved \citep{juodzbalis2026}, contribute $\lesssim10\%$ of the
H$\beta$ and H$\alpha$ flux.

To account for any residual total flux loss, the spectrum is then scaled to the NIRCam photometry with a single factor, defined as the median difference between the NIRCam photometry and the pseudo-photometry measured from the NIRSpec spectrum using the filter response curves.
The scale factors are listed in Table~\ref{tab:epochs}. We note that the UNCOVER photometry was taken before the DAJE spectroscopy, so for a variable source the scaling sets the continuum to its level at an earlier epoch. Because the factor does not depend on wavelength, it leaves the EWs,
the line widths and the line ratios unchanged; it enters only the
absolute fluxes, which we use for $L_{5100}$
(Section~\ref{sec:broad}) and in Figure~\ref{fig:dmag}, where its
effect is shown.
We assume the line spread function (LSF) from the latest calibration by \citet{shajib2025}.

The gap between the two NIRSpec detectors falls at different wavelengths in each
QSO1 image, although this is mitigated by wavelength overlap between the three gratings. Once joined, only one gap remains per image: 1.229--1.331\,$\mu$m (rest frame 1529--1656\,\AA) in image B (which removes C\,\textsc{iv}\,$\lambda$1549 and He\,\textsc{ii}\,$\lambda$1640), and 0.980--1.082\,$\mu$m (1219--1346\,\AA) in image C (which removes part of Ly$\alpha$). All other lines used here are covered in both images, which combined cover 1190--6840\,\AA\ in the rest frame without gaps.\footnote{Two faint emission-line galaxies, at $z=6.584$ and 8.597, observed in shutters 70--80\arcsec\ away, are dispersed onto the detector rows of image B. Their continua are undetected, and their strongest lines, H$\beta$, [O\,\textsc{iii}] and H$\alpha$, are not detected where they fall on the 2-D spectrum of B. The cluster member $0\farcs9$ from image B lies outside the shutters. A forward model of the contamination is deferred to future work.}

\section{Lens model and flux-ratio anomalies}
\label{sec:lensing}

The foreground cluster A2744 produces three images of QSO1, observed at three different epochs. Correcting the fluxes and correcting for gravitational time delay requires a gravitational lens model. 
Section~\ref{sec:epochs} summarizes the inferred magnifications and time delays from our new ``macro" model of the cluster (Di Rosa et al. 2026, in prep) and compares them with those from previous models. 
Section~\ref{sec:anomaly} shows that image A is brighter than
predicted by the macro models, and attributes the anomaly to millilensing, i.e., lensing by small
halos along the line of sight and in the cluster that the model does
not include. Section~\ref{sec:microlensing} uses detailed modeling to show that stellar microlensing is negligible, while millilensing can produce the observed anomaly. Section~\ref{sec:lenstests} uses spectroscopy to further confirm our hypothesis.

\subsection{Macro Model and Source-frame Epochs}
\label{sec:epochs}

We adopt version~1.9 (hereafter v1.9) of the strong lensing model of Abell~2744 (G. Di Rosa et al., in prep.), realized with \textsc{Gravity.jl} \citep{lombardi2024} following the mass parametrization of \citet{bergamini2023}. The model includes four cluster-scale haloes and one component per cluster member, whose masses follow scaling relations calibrated on measured stellar velocity dispersions \citep{bergamini2019, granata2026}. The model is constrained by the positions of 178 multiple images of 62 sources at $z=1.03$--9.76, all but three with spectroscopic redshifts. The image positions are reproduced with a collective r.m.s. residual of $0\farcs33$. The positions of the three images of QSO1 have an r.m.s. residual of $0.34''$. These are standard residuals for state of the art models that do not include substructure along the line of sight and in the plane of the lens. 

We report here the magnifications, the
delays, and their uncertainties, estimated as the 16th, 50th and 84th percentiles over 1000 realizations drawn from the MCMC chains of the model. The
magnifications are $\mu=5.98^{+0.20}_{-0.18}$,
$6.88^{+0.38}_{-0.39}$ and $3.30^{+0.09}_{-0.08}$ for images
A, B and C, respectively, and are computed at the model-predicted positions. The magnifications are stable across our \textsc{Gravity.jl} models. The systematic uncertainty, estimated from the spread between the models, is comparable to the statistical one.\footnote{A variant of the model without the galaxy close to image B gives $5.6\pm0.2$, $7.2^{+0.6}_{-0.4}$ and $3.1\pm0.1$. For comparison, the UNCOVER model of \citet{furtak2025} gives $5.4\pm1.1$, $7.5\pm1.5$ and $2.7\pm0.5$.} 
\par

Figure~\ref{fig:showpiece} shows the three images of QSO1 in the
NIRCam imaging of Abell~2744, with the critical curve of model~v1.9
at the redshift of QSO1, the MSA shutters of the spectra used here,
and the arrival delay of each image.
Our macro lens model gives arrival-time delays of
$\Delta t_{\rm AB}=2.28^{+0.08}_{-0.07}$\,yr between images A and B 
and $\Delta t_{\rm CA}=16.99^{+0.36}_{-0.40}$\,yr between C and A in 
the observed frame, and hence $\Delta t_{\rm BC}\simeq19.3$\,yr 
between B and C, or 2.4\,yr in the rest frame, the interval sampled 
by the DAJE spectra of the two images\footnote{The UNCOVER model of 
\citet{furtak2025} gives $2.34^{+0.20}_{-0.02}$\,yr and 
$19.28^{+0.64}_{-0.41}$\,yr for $\Delta t_{\rm AB}$ and 
$\Delta t_{\rm CA}$; the two models differ by $\simeq2.3$\,yr on the 
longer delay, i.e.\ by $\simeq0.29$\,yr in the rest frame. This 12\% 
difference is statistically significant but does not affect our 
conclusions.}.

The arrival order of the images is such that at a given observer time, B corresponds to the earliest rest-frame time, and C to the latest one. We express all epochs as source-frame times
$t_{\rm s}=(t_{\rm obs}-\Delta t_i-t_0)/(1+z_{\rm sys})$, where $t_0$ is the
date of the first UNCOVER observation of image A, $t_{\rm obs}$ the date of
observation and $\Delta t_i$ the delay of image $i$ ($\Delta t_{\rm A}=0$).
Table~\ref{tab:epochs} lists the seven spectra used here and their
rest-frame times in days.

\begin{deluxetable}{lcccccc}
\tablecaption{Spectra and Source-frame Epochs\label{tab:epochs}}
\tablehead{\colhead{Spectrum} & \colhead{Date (MJD)} & \colhead{Image} &
\colhead{$t_{\rm s}$ (d)} & \colhead{Scale} &
\colhead{$C_{\rm H\beta}$} & \colhead{$C_{\rm H\alpha}$}}
\startdata
UNCOVER PRISM & 60157.8 & A & 0 & 1.19 & 1.53 & 1.58\\
UNCOVER PRISM & 60157.6 & B & $-104$ & 0.72 & 1.57 & 1.55\\
UNCOVER PRISM & 60157.8 & C & 772 & 0.96 & 1.57 & 1.54\\
UNCOVER PRISM & 60522.3 & A & 45 & 1.11 & 1.48 & 1.62\\
SPURS G140M, G395M & 60988.6 & B & 0 & 0.86 & 1.38 & 1.48\\
DAJE G140M, G235M, G395M & 61364.5 & B & 46 & 0.77 & 1.53 & 1.41\\
DAJE G140M, G235M, G395M & 61364.5 & C & 922 & 0.86 & 1.85 & 1.88\\
\enddata
\tablecomments{$t_{\rm s}$ is the source-frame time relative to the
first UNCOVER observation of image A, computed with the delays of
model~v1.9 (Section~\ref{sec:epochs}). The SPURS and DAJE dates are
the exposure-weighted MJDs for G395M. The DAJE spectrum of image A is
contaminated and is not used (Section~\ref{sec:data}). Scale is the
factor by which the pipeline multiplies each spectrum to match the NIRCam
photometry. $C_{\rm H\beta}$ and $C_{\rm H\alpha}$ are the slit-loss
corrections at the observed wavelengths of H$\beta$ and H$\alpha$,
relative to a centered point source (Section~\ref{sec:data}).}
\end{deluxetable}

\subsection{The Flux of Image A Is Anomalous Due to Millilensing}
\label{sec:anomaly}

%%%%%%%% Figure: Delta m %%%%%%%%%%%%%%%%%%%%%%%%%%%%%%%%%%%%%%%%%%%
\begin{figure*}
\centering
\includegraphics[width=\textwidth]{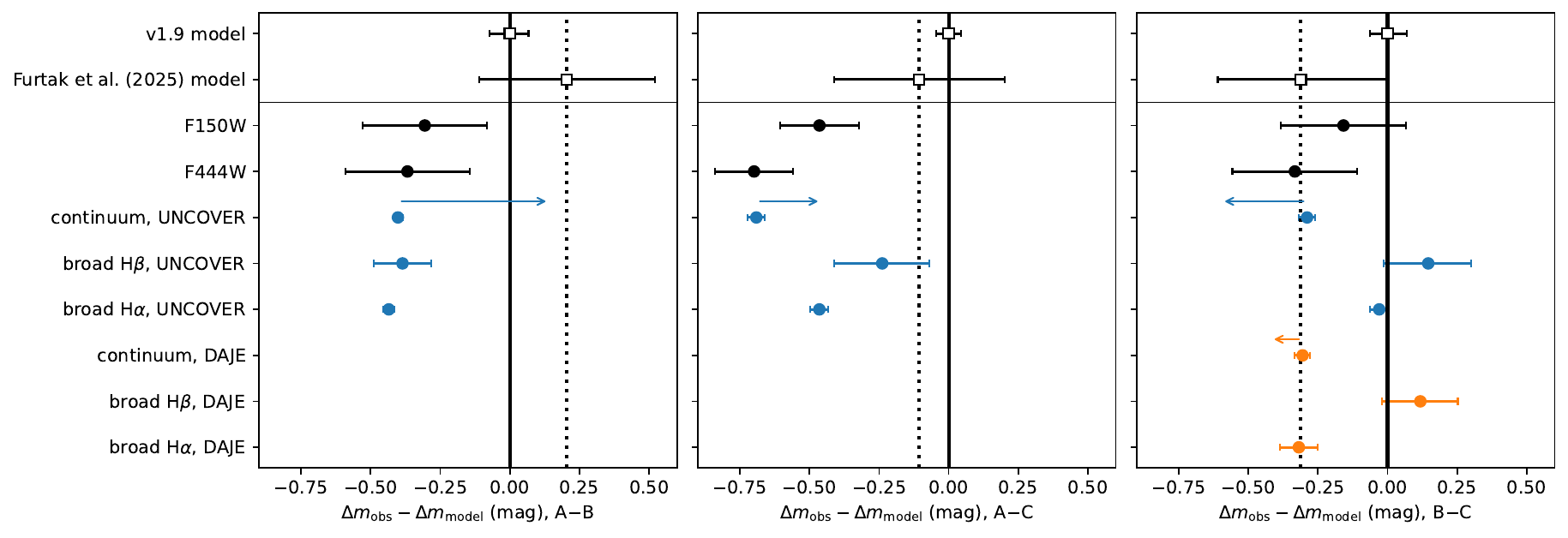}
\caption{Observed minus model-predicted magnitude differences of pairs of images of QSO1, $\Delta m_{\rm obs}-\Delta m_{\rm model}$, for the pairs A$-$B (left), A$-$C (middle) and B$-$C (right). Black points are the F150W and F444W photometry (rest frame $\simeq1900$ and $\simeq5500$~\AA; Section~\ref{sec:anomaly}). Blue points are the continuum at rest-frame 5100~\AA\ and the broad H$\beta$ and H$\alpha$ fluxes from the UNCOVER PRISM spectra; orange points are the same quantities from the DAJE-JWST spectra of images B and C. Each spectrum is normalized to the photometry by a single factor (Section~\ref{sec:data}). An arrow shows by how much all the points of that set of spectra move when this normalization is removed. The points use the magnifications of lensing model v1.9 (thick solid line at zero), and the dashed lines show where zero would lie if the magnifications of the
UNCOVER model of \citet{furtak2025} were used instead; their uncertainties are also plotted at the top.}
\label{fig:dmag}
\end{figure*}

The flux ratios of the images do not match the magnification ratios of the macro model. In the UNCOVER catalog \citep{weaver2024}, image B is flagged as contaminated by the nearby cluster member, and a separate object is blended into image C at the shorter wavelengths. To ensure reliable photometry, we directly measure the F150W and F444W fluxes with \textsc{galfit} on both the GLASS-JWST \citep{treu2022} and UNCOVER imaging data. Possible systematic differences between the two data sets stem from intracluster light subtraction. We adopt the mean of the measurements based on GLASS-JWST and UNCOVER as our fiducial value. We estimate the total error including systematics to be 0.1 mag for A and C and 0.2 mag for B due to its bright nearby galaxy (Figure~\ref{fig:showpiece} cutouts). In F150W, image A is brighter than predicted by 0.31\,mag relative to B and 0.46\,mag relative to C, while the difference between B and C agrees with the model to within
0.16\,mag, and to within 0.33\,mag in F444W ($1.5\sigma$ for the
uncertainties of 0.2 and 0.1\,mag of B and C; Figure~\ref{fig:dmag}).
The tension is present in all our \textsc{Gravity.jl} models, with and without the cluster member near B. 

Variability of the source does not provide a likely explanation for this pattern. Images A and B sample the source 104\,d apart and images B and C 876\,d apart. Therefore, a variable source would be more likely to produce a larger difference between B and C, the pair that agrees with the model. We thus conclude that the excess (or ``anomaly" in the strong lensing jargon) is specific to image A. We show below that it is consistent with millilensing, the effect of mass along the line of sight and of subhalos in the plane of the lens that are fainter than the faintest cluster members included in the macro model (F150W\,$=22$\,mag, $\sigma\simeq50$\kms).
Millilensing affects equally all sources with angular size smaller than milliarcseconds \citep{Treu2010}. It is thus achromatic for the sources considered here. Indeed, this is what we observe in Figure~\ref{fig:dmag}.

In contrast, an anomaly due to microlensing would be larger in the
continuum than in the more extended broad lines, absent in the narrow
lines, and larger in the rest-frame ultraviolet than in the optical,
which arises from a larger region. Instead, the anomaly of image A is
present in all bands, including the reddest, F444W.

\subsection{Microlensing and Millilensing}
\label{sec:microlensing}

Detailed modeling supports our conclusion that the anomaly of A is due to millilensing and not to microlensing. 
We estimate the inferred microlensing signal at the three image positions for the continuum, the broad-line region, and the narrow-line region, as described in Appendix~\ref{append:lensing}. Images A and C are virtually unaffected, while B has a small but non zero probability to be demagnified by 10--20\%. This is mostly because
image B lies closest to the cluster galaxies, so its stellar convergence is
over ten times that of A and C (Appendix~\ref{append:lensing}), and, as a
saddle-point image, it is the most sensitive to perturbations by stars.  The probability of a $\approx0.3-0.4$\,mag excess in A with a better consistency between B and C is $\lesssim0.1\%$, and thus microlensing cannot explain the observed anomaly.

We estimate the millilensing signal expected in CDM by forward modeling the populations of cluster subhalos and of halos along the line of sight with \texttt{pyHalo} \citep{Gilman++20} in Appendix~\ref{append:lensing}. At the source redshift of 7, halos along the line of sight dominate: alone, they perturb the image magnifications by $\sim5$--$9\%$ (the standard deviation of $\ln\mu$ across the three images), and with cluster subhalos of projected mass fraction $f_{\rm sub}\approx0.4$--$12\%$ the perturbations are $6$--$18\%$.
Propagated onto the observed magnitude differences, with the scatter of the two images of each pair added in quadrature to the photometric errors, these perturbations reduce the F150W excess of image A to $1.2\sigma$ (A$-$B) and $2.4\sigma$ (A$-$C) for $f_{\rm sub}\lesssim1\%$, and to $0.9\sigma$ and $1.7\sigma$ for $f_{\rm sub}=12\%$, while B$-$C remains consistent with the model in all cases. Millilensing can therefore account for the observed flux-ratio anomaly. Millilensing magnifies the continuum and the lines of an image alike,
so it leaves the EWs unchanged, while microlensing would change them.
We can therefore use the EWs to study variability. Millilensing also
does not change the time delays significantly for the purpose of this
study \citep{gilman2020b}.

An additional possibility is that the anomaly is the net outcome of B being microlensed while C is affected by a close companion. Image C has a faint compact galaxy $\sim0.25''$ away, which mainly shows up at short wavelengths. The companion galaxy, which is fainter than C at F150W and becomes negligible at F444W, is included in our image fitting, and thus unlikely to significantly affect our photometry, but its lensing impact is unknown.

\subsection{A spectroscopic test of microlensing}
\label{sec:lenstests}

Spectra of two images taken at the same source-frame time test 
microlensing independently of any intrinsic variability of QSO1, 
which we examine over all epochs in Section~\ref{sec:var}. An 
intrinsic change would appear in each image at a time set by its 
delay, while microlensing would change each image independently, 
over a multi-year timescale \citep{munoz2022, forestoribio2024}.

In our lens model, two pairs of spectra sample the source at nearly the same source-frame time: UNCOVER A1 and SPURS B, at $t_{\rm s}=0$ and 0\,d, and UNCOVER A2 and DAJE B, at 45 and 46\,d (Table~\ref{tab:epochs}). For each pair of images, a difference in EW  
would be the ``smoking gun" of differential microlensing of A and B (modulo systematic differences between the PRISM and grating spectra). The H$\beta$ EWs are $45_{-4}^{+4}$ and $41_{-6}^{+6}$\,\AA\ in the first pair and $56_{-6}^{+7}$ and $44_{-2}^{+2}$\,\AA\ in the second, consistent within the uncertainties (Table~\ref{tab:lines}). However, the H$\alpha$ EWs of image A are systematically smaller than those of image B by $\approx11$--16\%. This offset is comparable to the systematic uncertainty of the EWs set by
the choice of continuum window and line model ($\approx5$\% for H$\beta$
and $\approx10$\% for H$\alpha$; Appendix~\ref{append:compare_fit}). Therefore, we conclude that
we do not detect differential
microlensing between images A and B.

The narrow lines would measure the magnification ratios including
millilensing \citep{paugnat2026}. [O\,\textsc{iii}]\,$\lambda5007$ is
detected at $>3\sigma$ in UNCOVER A1, SPURS B and both DAJE spectra,
but its flux ratios between images carry the differential slit losses
of up to 25\% (Section~\ref{sec:data}), so they do not test the
magnification ratios independently. Within each spectrum, however, the
slit losses and the magnification cancel, and we use
[O\,\textsc{iii}]\,$\lambda5007$ to normalize the line and continuum
fluxes in Section~\ref{sec:var}.

\section{Spectral analysis}
\label{sec:analysis}

In this section we present measurements of each image spectrum. Section~\ref{sec:fitting} describes the line fitting and the systematic
uncertainties set by the choice of model and continuum.  Section~\ref{sec:var} demonstrates that the H$\beta$ and H$\alpha$ EWs do not change significantly over the 2.8 rest-frame yr sampled by the seven spectra, once uncertainties are taken into account.
Section~\ref{sec:broad} measures the widths of the broad Balmer lines
and uses them to estimate black-hole mass. Section~\ref{sec:heii} analyzes the helium
emission lines, which probe the gas density. 
Section~\ref{sec:decrement} analyzes the Balmer decrement
to constrain reddening due to dust or high density gas.
Section~\ref{sec:metal} uses narrow
lines, to show that the metallicity of the gas is low.

\subsection{Line Fitting}
\label{sec:fitting}

\begin{figure*}
\centering
\includegraphics[width=\textwidth]{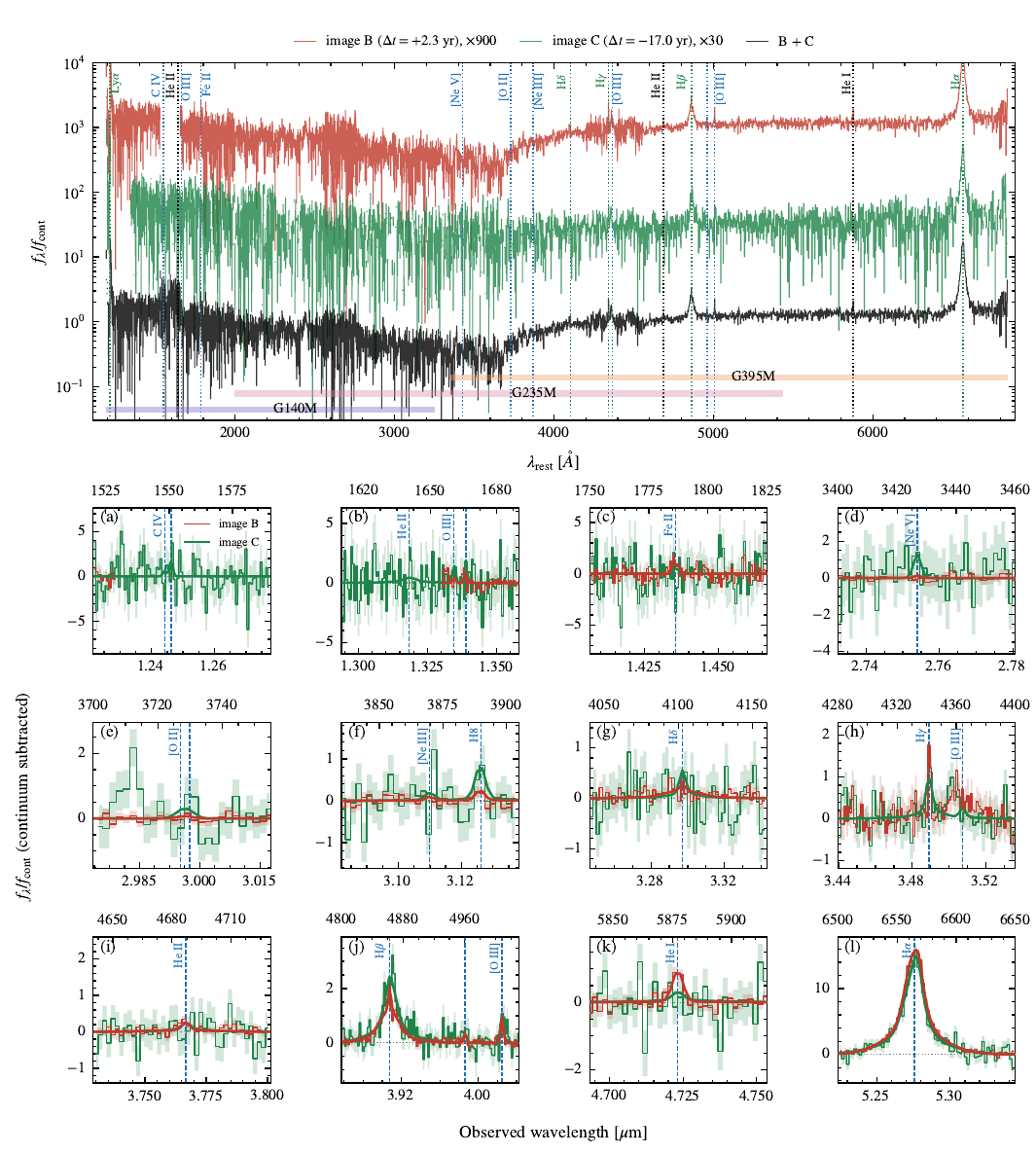}
\caption{DAJE-JWST NIRSpec spectra of images B and C of QSO1 and the
emission lines used in this paper. Top: the spectra of images B (red; $\times900$) and C (green; $\times30$) and their combination (black), with the lines of the lower panels marked; the legend gives the time delay of each image relative to image A. Lower panels: the lines in the spectra of images
B (red) and C (green). Each spectrum is divided by its median continuum at 3.30--3.60\,$\mu$m. The local continuum is subtracted in each window. Shaded bands are the $1\sigma$ uncertainties, smooth curves
the best-fitting models, and dashed lines mark the lines at $z_{\rm sys}$. The bottom horizontal axes give the observed wavelength. The top horizontal axes give the rest-frame wavelength.}
\label{fig:lines}
\end{figure*}

Figure~\ref{fig:lines} shows wavelength zooms of all the lines we fit in the DAJE spectra of
images B and C. All spectra are fitted with \texttt{unite} \citep{hviding_unite}. The velocity $v$ is measured relative to $z_{\rm sys}=7.0364$.
Each line is fitted independently. 
For H$\alpha$ we use a wider fitting window than for the other lines: the rest-frame interval 5300--6700~\AA, gives sufficient coverage and signal to noise ratio to constrain the continuum. This choice is similar to that of \citet{furtak2025}, who adopted a global continuum. Appendix~\ref{append:compare_fit} quantifies how the results depend on the fitting window. The dependency on the fitting window is taken into account in our estimation of systematic errors.

%%%%%%%% Figure 2 %%%%%%%%%%%%%%%%%%%%%%%%%%%%%%%%%%%%%%%%%%%

In the DAJE spectra, two broad Gaussians in addition to a narrow component (also assumed to be Gaussian) fit H$\beta$ slightly better than one, although a single Gaussian remains an acceptable fit. We use two, the
same model as for H$\alpha$. Adopting a single Gaussian would have changed the H$\beta$ EW and flux by
$\lesssim10\%$ (Appendix~\ref{append:compare_fit}). For H$\alpha$, a single broad Gaussian
leaves a residual component of ${\rm FWHM}\simeq700$\,\kms\ (Section~\ref{sec:broad}),
which we model in two ways: as part of the broad profile, with the broad
component described by two Gaussians (as in \citealt{deugenio2025}), or as a
separate intermediate component with its own width. Other broad profiles, such as an exponential, give comparable fits but are
less favored physically \citep[][Appendix~A]{deugenio2025}.
\texttt{unite} integrates the model over each pixel, which accounts for the
undersampling of the LSF, and convolves it with the LSF of
\citet{shajib2025} (Section~\ref{sec:data}). The fitted widths are
therefore those of the intrinsic profile.

DAJE's image B spectrum shows evidence of absorption on top of the
H$\beta$ emission, which we model with a Gaussian optical depth applied to
the broad component and the continuum. The absorption is statistically
required ($\Delta\chi^2=66$ for three additional parameters,
$\approx7.6\sigma$), and its fitted parameters are consistent with the
higher-resolution BlackTHUNDER measurements
(Appendix~\ref{append:comparison}; Figure~\ref{fig:absorbHb}). We do not
include the absorption component hereafter, unless otherwise noted, and
focus on the total EW of the whole line, which is measured directly and
does not depend on the decomposition into emission and absorption.

\begin{figure*}
\centering
\includegraphics[width=0.8\linewidth]{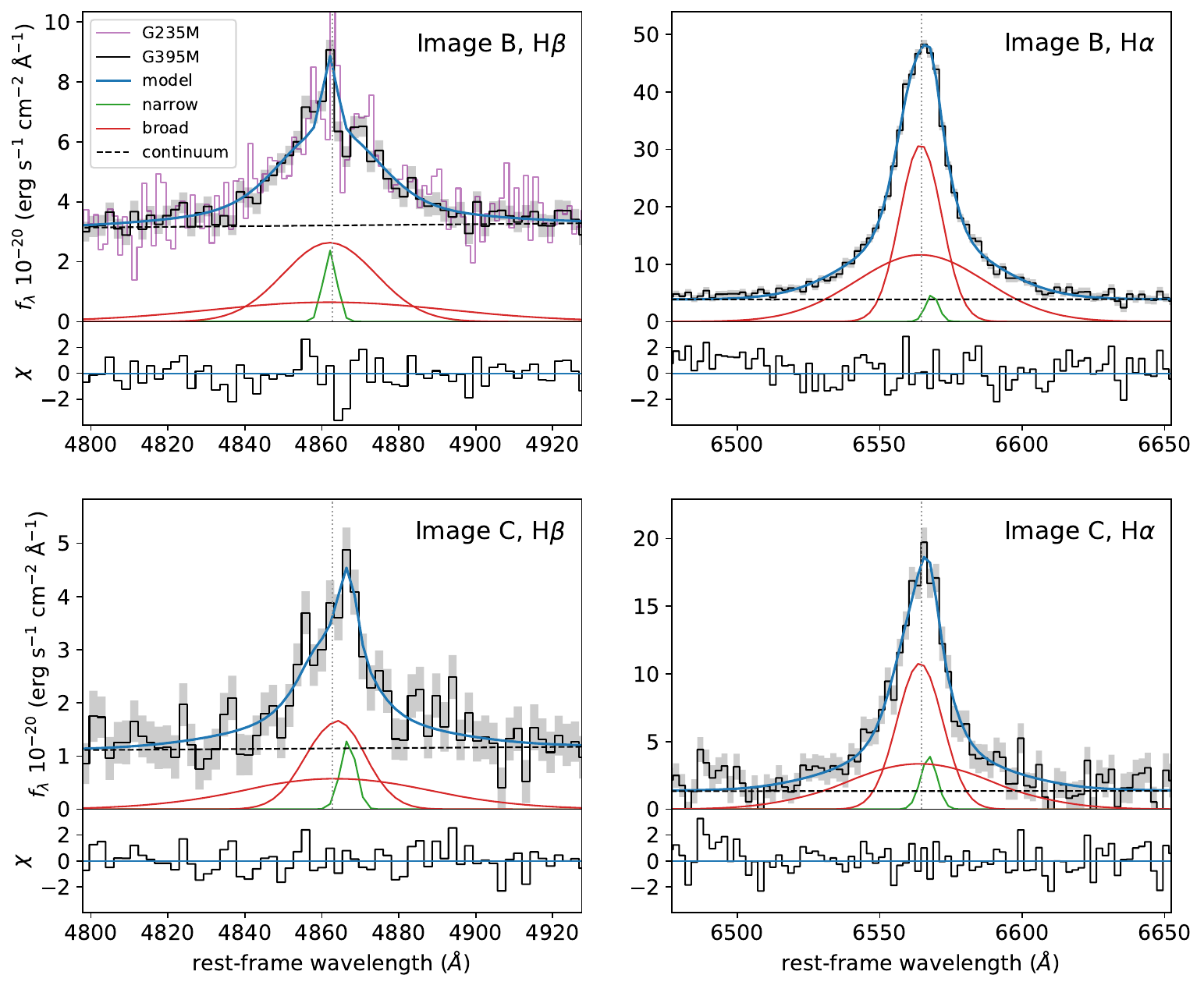}
\caption{Fits to H$\beta$ (left) and H$\alpha$ (right) profiles in the DAJE spectra of images B (top) and C (bottom). In each panel, the black line represents the G395M spectrum  and the gray band its $1\sigma$ uncertainty. H$\beta$ in image B also has G235M coverage, plotted as purple, which is fitted jointly with G395M. The flux densities are in the rest frame and not corrected for magnification. The blue curve is the total model shown at the resolution of G395M, which is the sum of the narrow component (green), the broad component (red), and the continuum (dashed). The lower panels show the residuals in units of the uncertainty.}
\label{fig:fitbalmer}
\end{figure*}

For the grating data (DAJE and SPURS), all other line parameters are freely varied in the fitting. We show the observed H$\alpha$ and H$\beta$ lines for DAJE images B and C and best fits in Figure~\ref{fig:fitbalmer}. For the
UNCOVER PRISM data the multi-Gaussian model is degenerate. To reduce the
degeneracy and keep the EWs comparable with the grating data, we fix the
width of the narrow component, the
width ratio of the two broad Gaussians, and the flux ratios and velocity
offsets between the components to the values measured in the DAJE spectrum
of the same image. For the two UNCOVER spectra of image A we use the profile of DAJE image B, which samples the source within a day of the second UNCOVER epoch of A (Table~\ref{tab:epochs}). We have confirmed that the fitted results do not materially change if adopting the profile of image C. The fitting parameters for each line are then the overall
amplitude, a single scaling factor applied to the widths of both broad
Gaussians, and a common velocity shift. Appendix~\ref{append:comparison} compares these fits with the published
measurements.
The EWs depend on the line model and on the wavelength range used to
define the continuum, and this dependence differs between spectra: for
H$\beta$ the EW changes by $\approx5$\%, and for H$\alpha$, which lies at
the red end of G395M, by $\approx10$\% (Appendix~\ref{append:compare_fit}). We adopt
these differences as systematic uncertainties. Independent fits of two
reductions of the DAJE data give widths, fluxes and EWs consistent within
the uncertainties. The larger uncertainties of image C reflect its
continuum signal-to-noise ratio, about five times lower than that of B.
Narrow-line ratios are uncorrected for reddening because \citet{deugenio2025} showed that the narrow-line dust attenuation is consistent with 0 ($A_V=-0.2\pm0.5$\,mag). For the broad-component ratio of He\,\textsc{i}~$\lambda$5876/H$\beta$, we quote both the observed ratio and the
ratio corrected with the $A_V$ of Section~\ref{sec:decrement}, because the broad
Balmer decrement may not measure dust. Ratios of lines within $\sim$200\,\AA,
such as $R_3$ and He\,\textsc{ii}~$\lambda$4686/H$\beta$, change by at most $\simeq$
0.05\,dex for $A_V=3$ and the SMC extinction curve \citep{gordon2003}, and we leave them
uncorrected.

\subsection{Variability}
\label{sec:var}
We test whether QSO1 varies by comparing the H$\beta$ and H$\alpha$ EWs of
the seven spectra of Table~\ref{tab:epochs}, which sample the source over
2.8 rest-frame yr. We focus on EWs because they are independent of slit-losses, flux calibration, and achromatic lensing magnification.
We measure the intrinsic rms variability \citep{Shen19}
and its significance following \citet{Zou26}. The EWs of each line are
described as $x_i\sim\mathcal{N}(m,\sigma_0^2+\sigma_i^2)$, where $x_i$ is the
EW of spectrum $i$, $\sigma_i$ its statistical uncertainty combined in
quadrature with the systematic uncertainty of the line fitting (the spread
of the EW over the fitting choices of Appendix~\ref{append:compare_fit}: $\approx5\%$
for H$\beta$ and $\approx10\%$ for H$\alpha$), and $m$ and $\sigma_0$ are free
parameters. The significance is computed by comparing the maximum likelihood
of this model with that of a constant source ($\sigma_0=0$), and
$\sigma_0/m$ is the fractional variability, for which we quote 95\% upper
limits. Without the systematic term, H$\beta$ still shows no significant
variability, and neither does H$\alpha$ when the continuum is fitted in the shorter windows of
Appendix~\ref{append:compare_fit}.

Figure~\ref{fig:ew} shows the EWs against source-frame time. Neither line
varies significantly: the significance is $1.1\sigma$ and $0\sigma$ for
H$\beta$ and H$\alpha$, and $\sigma_0/m<32\%$
and $<15\%$, respectively. Excluding image A gives similar results. Between images B and C, which sample the source
2.4 rest-frame yr apart in the DAJE spectra, the EWs differ by 54\%
in H$\beta$ and 7\% in H$\alpha$, with significance of $2.8\sigma$ and $0.4\sigma$ respectively. The
H$\beta$ EWs of image C, 72 and 67\,\AA\ in the UNCOVER and DAJE spectra,
exceed those of the five spectra at $t_{\rm s}\le46$\,d, 41--56\,\AA, by
$52\pm16$\% ($3.5\sigma$). This excess also does not contradict the non-variability based on the $\sigma_0$ model. The latter describes a scatter common to all epochs and is weighted toward the most precise spectra, which agree with each other, whereas the excess is confined to the two spectra with the lowest signal-to-noise ratios. It is a difference between the early and the late epochs, which the gap in the time coverage separates, and not evidence for stochastic variability. Since the only spectra after
$t_{\rm s}\simeq50$\,d are of image C, a change of the source between 50
and 770\,d cannot be separated from a possible systematic specific to image C.

We also assess the flux variability for H$\alpha$, H$\beta$, and the rest-frame 5100\,\AA\ continuum. Their comparison between images
depends on the magnifications, which carry the flux-ratio anomaly of image A
(Section~\ref{sec:anomaly}) and the spread of $\mu_{\rm B}/\mu_{\rm C}$ among
the lens models, and on slit-loss differences of up to 25\% between B and C
(Section~\ref{sec:data}). To mitigate these effects, we instead normalize them by the [\iona{O}{iii}] $\lambda5007$ flux, which is not expected to vary. After the normalization, neither the continuum nor the broad-line fluxes vary at a statistically significant level ($<2\sigma$). $\sigma_0/m$ is $<50$\%, 47\%, and 34\% for the continuum, broad H$\alpha$ flux, and broad H$\beta$ flux, respectively, and these limits include the statistical uncertainty of [\iona{O}{iii}] $\lambda5007$ itself.

The published evidence for variability in QSO1 rests on EW differences
between images C and A in the UNCOVER PRISM spectra
(Section~\ref{sec:intro}). Image A is also the image whose flux departs
from the lens model (Section~\ref{sec:anomaly}), but the anomaly can
change the EWs only if it acts differently on the continuum and on the
broad lines, as microlensing does. Millilensing and errors of the model
magnify continuum and lines alike. Microlensing of A is unlikely
(Section~\ref{sec:microlensing}), and the two pairs of spectra that sample
the source at the same time show no EW difference between A and B beyond
the systematic uncertainty (Section~\ref{sec:lenstests}). The changes
reported by \citet{furtak2025}, 22\% in H$\beta$ and 18\% in H$\alpha$
relative to image C, carry statistical errors only; our fits of the same
spectra give 37\% and 25\% (Table~\ref{tab:lines}), within $2.2\sigma$
once the systematic uncertainty is included. The larger change found by
\citet{ji2025} depends on the extraction aperture (Section~\ref{sec:intro}),
and their limit of 45 light-days on the size of the broad-line region
assumes that the differences are intrinsic.

LRDs as a population vary
less than blue broad-line AGN of the same luminosity: the H$\alpha$ rms is
$\lesssim4$\% over 1--3 rest-frame months in NEXUS \citep{stone2026}, and the
$1\sigma$ limits are 5--30\% over 72--187\,d in OCEANS \citep{brooks2026}. An average SDSS quasar varies by $\simeq0.2$\,mag between two epochs over our
baseline \citep{macleod2010}, i.e.\ $\simeq13$\% in the continuum. Over an
interval longer than the light-crossing time of the broad-line region, the
EW changes by $(1-\eta)$ times the continuum, where $\eta\simeq0.5$--1 is the
responsivity of the line \citep{korista2004}, so an ordinary quasar would
change the EWs by $\lesssim7$\%, below the uncertainty of each EW. 
The EWs therefore do not distinguish QSO1 from an ordinary quasar, 
and the [O\,\textsc{iii}]-normalized flux limits (35--50\%) likewise 
exceed the $\simeq13\%$ continuum change of an average quasar. In 
summary, the variability previously reported for QSO1 is not 
confirmed, but the current limits are a factor of $\approx2$ too loose 
to separate a sub-Eddington disk, which should vary as quasars do, 
from the envelope and wind models, which predict slower variations.

\subsection{Broad Balmer Lines}
\label{sec:broad}

Microlensing magnifies the high-velocity component, which arises closest to
the black hole. Since the low-velocity part of a line is expected to form further out \citep{paugnat2026}, the
widths can differ between images for reasons unrelated to the source. We
therefore compare H$\beta$ and H$\alpha$ within each image. The two lines
fall in the same grating, so their ratios do not depend on the
magnification or the slit losses.

Table~\ref{tab:lines} lists the line measurements. The H$\alpha$ profile
contains an intermediate component of ${\rm FWHM}\simeq700$\,\kms\ ($762^{+26}_{-25}$\,\kms\
in image B) with no counterpart of similar width in H$\beta$; if it
has none, this component has H$\alpha$/H$\beta>28$ ($3\sigma$;
Section~\ref{sec:decrement}). We adopt as our fiducial model the decomposition in
which this component is distinct from the broad line:
the H$\alpha$ profile requires it statistically, and
its width, a third of that of the broad line,
identifies gas at lower velocity -- at larger radii
than the broad-line region if the motion is virial --
whereas an H$\beta$ counterpart is not required by the
data, although it is not excluded
(Section~\ref{sec:decrement}). For comparison we also
quote the fits with the component included in the broad
profile. If the intermediate component is included in the
broad profile, the FWHMs of the broad H$\beta$ and H$\alpha$ in image B are
$1969^{+116}_{-113}$ and $935^{+20}_{-19}$\,\kms, and the ratio of the
H$\beta$ to H$\alpha$ widths is $2.1\pm0.1$ and
$1.4\pm0.4$ in B and C (weighted mean $2.0\pm0.1$). The
difference lies in the narrow Gaussians of the two profiles (width ratio $2.1\pm0.2$ in
image B) and less in the wide Gaussians ($1.8\pm0.2$); measured on the profiles as a whole, the full width of H$\beta$ is 2.1--2.2 times that of H$\alpha$ at 80\% and 50\% of the peak, and 1.6 times at 10\%. If the intermediate component is fitted separately, broad H$\alpha$ is the
remaining emission: the H$\beta$ FWHM is $0.8\pm0.1$ times that of broad H$\alpha$ and its line dispersion (the second moment of the profile) $1.2\pm0.1$ times; the two measures lie on either side of unity, consistent with comparable widths.

If the intermediate component has no counterpart in
H$\beta$, its Balmer decrement exceeds that of the
whole broad line by a factor of $>3$
(Section~\ref{sec:decrement}), so it would be emitted
by gas in different physical conditions from the rest
of the line. With it fitted separately, broad H$\beta$
and broad H$\alpha$ have comparable widths (FWHM ratio
$0.8\pm0.1$).
%%%%%%%% Figure 3 %%%%%%%%%%%%%%%%%%%%%%%%%%%%%%%%%%%%%%%%%%%
\begin{figure}
\centering
\includegraphics[width=\columnwidth]{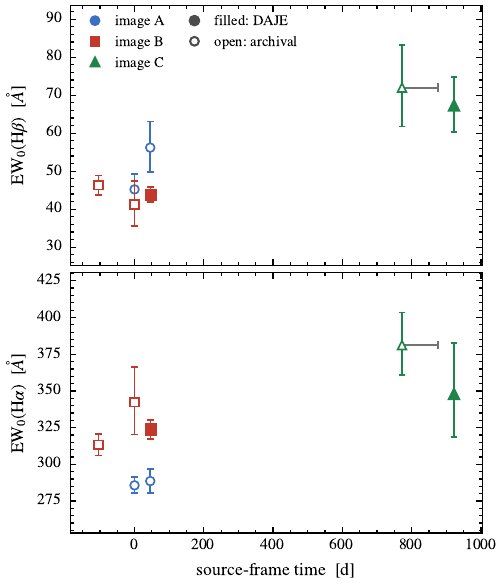}
\caption{Rest-frame equivalent width of the H$\beta$ (top) and
H$\alpha$ (bottom) lines against source-frame time
(Section~\ref{sec:epochs}). Colors and symbols denote the image; filled
symbols are the DAJE gratings, open symbols the archival UNCOVER PRISM and
SPURS G395M spectra (Table~\ref{tab:epochs}). The gray bar on the archival
point of image C spans the shift of that point between the adopted lens model
and that of \citet{furtak2025}. The EWs have $\approx5\%$ and $\approx10\%$ systematic uncertainties from fitting settings (Appendix~\ref{append:compare_fit}).}
\label{fig:ew}
\end{figure}

\begin{deluxetable*}{lccccccc}
\tablecaption{Broad H$\beta$ and H$\alpha$\label{tab:lines}}
\tablehead{
\colhead{} & \multicolumn{4}{c}{UNCOVER} & \colhead{SPURS} & \multicolumn{2}{c}{DAJE} \\
\cline{2-5} \cline{7-8}
\colhead{} & \colhead{A1} & \colhead{A2} & \colhead{B} & \colhead{C} & \colhead{B} & \colhead{B} & \colhead{C}}
\startdata
EW(H$\beta$) (\AA) & $45_{-4}^{+4}$ & $56_{-6}^{+7}$ & $46_{-3}^{+3}$ & $72_{-10}^{+11}$ & $41_{-6}^{+6}$ & $44_{-2}^{+2}$ & $67_{-7}^{+8}$ \\
EW(H$\alpha$) (\AA) & $286_{-5}^{+6}$ & $289_{-8}^{+8}$ & $313_{-7}^{+7}$ & $381_{-20}^{+22}$ & $342_{-22}^{+24}$ & $324_{-6}^{+6}$ & $348_{-29}^{+35}$ \\
H$\alpha$/H$\beta$ & $8.3_{-0.6}^{+0.7}$ & $9.1_{-0.9}^{+1.1}$ & $8.0_{-0.4}^{+0.4}$ & $6.8_{-0.8}^{+1.0}$ & $10.6_{-1.4}^{+1.7}$ & $9.4_{-0.4}^{+0.5}$ & $6.3_{-0.8}^{+0.9}$ \\
\hline
\multicolumn{8}{c}{H$\beta$ profile} \\
\hline
FWHM$_{\rm narrow}$ (\kms) & \nodata & \nodata & \nodata & \nodata & $1615_{-456}^{+495}$ & $1727_{-131}^{+140}$ & $1102_{-286}^{+289}$ \\
FWHM$_{\rm wide}$ (\kms) & \nodata & \nodata & \nodata & \nodata & $3685_{-1167}^{+941}$ & $4620_{-576}^{+279}$ & $3788_{-1000}^{+862}$ \\
$f_{\rm wide}$ & \nodata & \nodata & \nodata & \nodata & $0.48_{-0.28}^{+0.26}$ & $0.39_{-0.09}^{+0.08}$ & $0.54_{-0.16}^{+0.14}$ \\
width factor & $1.12_{-0.16}^{+0.18}$ & $1.17_{-0.24}^{+0.26}$ & $0.82_{-0.09}^{+0.09}$ & $1.69_{-0.42}^{+0.51}$ & \nodata & \nodata & \nodata \\
\hline
\multicolumn{8}{c}{H$\alpha$ profile} \\
\hline
FWHM$_{\rm narrow}$ (\kms) & \nodata & \nodata & \nodata & \nodata & $806_{-86}^{+94}$ & $762_{-25}^{+26}$ & $834_{-141}^{+161}$ \\
FWHM$_{\rm wide}$ (\kms) & \nodata & \nodata & \nodata & \nodata & $2440_{-296}^{+368}$ & $2436_{-110}^{+120}$ & $2799_{-567}^{+1225}$ \\
$f_{\rm wide}$ & \nodata & \nodata & \nodata & \nodata & $0.50_{-0.08}^{+0.07}$ & $0.54_{-0.02}^{+0.02}$ & $0.50_{-0.10}^{+0.10}$ \\
width factor & $1.05_{-0.02}^{+0.02}$ & $0.89_{-0.03}^{+0.03}$ & $0.95_{-0.02}^{+0.02}$ & $1.04_{-0.05}^{+0.06}$ & \nodata & \nodata & \nodata \\
\hline
FWHM(H$\beta$)/FWHM(H$\alpha$) & \nodata & \nodata & \nodata & \nodata & $1.97_{-0.43}^{+0.53}$ & $2.11_{-0.12}^{+0.13}$ & $1.35_{-0.34}^{+0.36}$ \\
\enddata
\tablecomments{EWs are rest-frame and include the narrow component and,
for H$\alpha$, the intermediate component. The broad profile is the sum
of a narrow and a wide Gaussian with a common centroid; $f_{\rm wide}$
is the fraction of the broad flux in the wide Gaussian. In H$\alpha$
the narrow Gaussian corresponds to the intermediate component
(Section~\ref{sec:broad}). For the UNCOVER PRISM A1, A2, and B spectra
the profile shape is fixed to that of the DAJE B spectrum, and for the
UNCOVER PRISM C spectrum to that of DAJE C; the only free shape
parameter is the width factor, which multiplies both Gaussian widths.
FWHM(H$\beta$)/FWHM(H$\alpha$) is the ratio of the FWHMs of the summed
broad profiles, and H$\alpha$/H$\beta$ the ratio of the broad fluxes,
both with the intermediate component of H$\alpha$ included; the values
with it fitted separately are given in Section~\ref{sec:broad}. The
helium lines, the higher Balmer lines and the narrow lines are in
Table~\ref{tab:other}. The DAJE spectrum of image A is not used
(Section~\ref{sec:data}).}
\end{deluxetable*}

We estimate the black-hole mass $M_\mathrm{BH}$ from the FWHM of broad
H$\beta$ and the continuum luminosity at 5100\,\AA, $L_{5100}$,
following \citet{shen2011}.
The luminosities are measured corrected for the magnifications and not corrected for extinction. We obtain $\log(M_\mathrm{BH}/M_\odot)=7.08\pm0.05$ and $6.69^{+0.18}_{-0.21}$ for images B and C (statistical uncertainties only), statistically consistent with each other. These values are 0.6--1\,dex below the dynamical mass, $\log(M_\mathrm{BH}/M_\odot)=7.7\pm0.3$ \citep{juodzbalis2026}, but consistent with it given the nominal 0.4-dex scatter of single-epoch estimators \citep{shen2013}. If the broad Balmer decrement is due to dust reddening ($A_V\simeq1.5-3$; Section~\ref{sec:decrement}), correcting $L_{5100}$ raises the masses by up to 0.7\,dex. In either case, $L_{5100}$ traces the ionizing luminosity only if the optical continuum comes directly from the accretion flow, which the MIRI photometry makes doubtful \citep{iani2026}. With the intermediate component fitted separately, broad H$\alpha$, under the relation in \citet{reinesvolonteri2015},
gives $\log(M_\mathrm{BH}/M_\odot)=7.28\pm0.04$ and
$7.33^{+0.29}_{-0.18}$ for images B and C; if the intermediate component were included in the broad profile,
the H$\alpha$ mass would lie 1.2\,dex below the dynamical mass. We discuss what these widths and masses imply for the origin of the
line broadening in Section~\ref{sec:widths}.

\subsection{\texorpdfstring{He\,\textsc{ii} $\lambda\lambda$1640, 4686 and He\,\textsc{i} $\lambda$3889, 5876}{He II 1640, 4686 and He I 5876}}
\label{sec:heii}

Every He$^{+}$ ion that recombines was ionized by a photon above 54\,eV, and
every He$^{0}$ atom by a photon above 24.6\,eV. The He\,\textsc{ii} and He\,\textsc{i} lines relative to H$\beta$ are related to how many of these photons are absorbed by the line-emitting gas (Section~\ref{sec:ionizing}).
He\,\textsc{ii} $\lambda$4686 falls at 3.767\,$\mu$m and He\,\textsc{i}
$\lambda$5876 at 4.723\,$\mu$m, both in G395M, and He\,\textsc{ii}
$\lambda$1640 at 1.318\,$\mu$m, in G140M. 
He\,\textsc{ii}\,$\lambda$1640 falls in the detector gap in image B (Section~\ref{sec:data}). For each line we fit the spectrum of each image, and of the SPURS spectrum of image B, with a narrow component with the profile of [\iona{O}{iii}] $\lambda$5007 and a broad component with the profile of broad H$\beta$ scaled in amplitude, and quote the
fluxes relative to narrow and broad H$\beta$, respectively. For He\,\textsc{i} $\lambda$5876 in DAJE B, which has a high SNR (see Table~\ref{tab:other}), we allow its narrow component to have a free width.

For a non-detection we report the 99.87th percentile of the posterior of the
amplitude (Table~\ref{tab:other}).
We obtain He\,\textsc{ii}/H$\beta<0.14$ in
image B, the deepest spectrum (Table~\ref{tab:other}).
He\,\textsc{i}\,$\lambda5876$ is detected in image B, at the position of QSO1 in the two-dimensional spectrum. The line is unresolved, with an intrinsic FWHM\,$<$\,396\kms. It is narrower than the intermediate-width component of FWHM\,$=560\pm50$\kms\ that \citet{deugenio2025} find in H$\alpha$, and we classify it as narrow; whether it arises in the narrow-line region or in denser gas is discussed in Section~\ref{sec:ionizing}. Its ratio to narrow H$\beta$ is $1.4_{-0.2}^{+0.3}$, or $0.5\pm0.1$ if narrow H$\beta$ is corrected for the absorption (Figure~\ref{fig:absorbHb}). He\,\textsc{i}\,$\lambda3889$ is undetected in image B: $\lambda3889/\lambda5876<0.5$, below the recombination value of 0.8 \citep{porter2012}. No broad He component is detected in the current data: broad He\,\textsc{i}/H$\beta<$\,0.13 (0.09 if corrected for $A_V=1.5$\,mag; Section~\ref{sec:decrement}), and its EW is below 5\,\AA, against $\simeq20$\,\AA\ typical of local LRDs \citep{park2026}. He\,\textsc{i} is not detected in image C, where the limit is consistent with the line in B given the lower magnification.
In the SPURS spectrum $\lambda5876$ falls in the gap between the 
detectors. The UNCOVER PRISM spectra show an excess at the 
wavelength of He\,{\sc i}\,$\lambda5876$, not reported previously, 
at $>3\sigma$ in image B and $\approx2\sigma$ in image A; the DAJE 
G395M data confirm this line.

\begin{deluxetable}{llcccc}
\tablecaption{DAJE Line Fluxes\label{tab:other}}
\tablehead{
\colhead{Line} & \colhead{Component} & \colhead{$\lambda_{\rm obs}$} & \colhead{B} & \colhead{C} \\
\colhead{} & \colhead{} & \colhead{($\mu$m)} & \colhead{} & \colhead{}}
\startdata
C\,\textsc{iv} $\lambda\lambda$1548,1551 & narrow & 1.245 & \nodata & $<14.3$ \\
He\,\textsc{ii} $\lambda$1640 & broad & 1.318 & \nodata & $<25.5$ \\
O\,\textsc{iii}] $\lambda\lambda$1661,1666 & narrow & 1.337 & $<13.6$ & $<14.2$ \\
Fe\,\textsc{ii} $\lambda$1786 & narrow & 1.436 & $10.6_{-2.9}^{+3.0}$ & $<11.3$ \\
{}[Ne\,\textsc{v}] $\lambda$3426 & narrow & 2.754 & $<4.8$ & $<12.2$ \\
{}[O\,\textsc{ii}] $\lambda\lambda$3727,3729 & narrow & 2.996 & $<4.6$ & $<8.6$ \\
{}[Ne\,\textsc{iii}] $\lambda$3869 & narrow & 3.110 & $<3.9$ & $<5.0$ \\
He\,\textsc{i} $\lambda$3889 & narrow & 3.126 & $<6.3$ & $<9.1$ \\
H$\delta$ & broad & 3.297 & $15.1_{-4.0}^{+3.8}$ & $<17.0$ \\
H$\delta$ & narrow & 3.297 & $<4.8$ & $<6.9$ \\
H$\gamma$ & broad & 3.489 & $<29.4$ & $<22.6$ \\
H$\gamma$ & narrow & 3.489 & $10.0_{-1.6}^{+1.7}$ & $<9.3$ \\
{}[O\,\textsc{iii}] $\lambda$4363 & narrow & 3.507 & $<7.7$ & $<4.8$ \\
He\,\textsc{ii} $\lambda$4686 & broad & 3.767 & $<17.5$ & $<15.2$ \\
He\,\textsc{ii} $\lambda$4686 & narrow & 3.767 & $<5.6$ & $<4.1$ \\
H$\beta$ & broad & 3.908 & $129.9_{-6.4}^{+6.3}$ & $69.5_{-7.8}^{+8.3}$ \\
H$\beta$ & narrow & 3.908 & $10.6_{-1.3}^{+1.3}$ & $<17.5$ \\
H$\beta$ (with absorption)$^a$ & narrow & 3.908 & $31.5_{-3.9}^{+3.4}$ & ... \\
{}[O\,\textsc{iii}] $\lambda$5007 & narrow & 4.025 & $11.9_{-1.0}^{+1.0}$ & $6.2_{-1.6}^{+1.7}$ \\
He\,\textsc{i} $\lambda$5876 & broad & 4.723 & $<16.7$ & $<19.5$ \\
He\,\textsc{i} $\lambda$5876 & narrow & 4.723 & $14.5_{-2.0}^{+2.1}$ & $<6.9$ \\
{}[Fe\,\textsc{vii}] $\lambda$6087 & narrow & 4.893 & $<4.8$ & $<6.9$ \\
H$\alpha$ & broad & 5.276 & $1228_{-16}^{+15}$ & $439_{-27}^{+28}$ \\
H$\alpha$ & narrow & 5.276 & $27.4_{-6.7}^{+8.6}$ & $<73$ \\
\enddata
\tablecomments{Observed fluxes in units of $10^{-20}$\,erg\,s$^{-1}$\,cm$^{-2}$,
not corrected for magnification; upper limits are the 99.87th percentile of the posterior for non-negative flux. Helium lines are fitted with a free-width component for image B and a broad component with the broad H$\beta$ profile (Section~\ref{sec:heii}). We regard the broad H$\delta$ detection in image B to be tentative as the significance is sensitive to the fitting window.\\
$^a$ The narrow emission flux if including a H$\beta$ absorption component in image B (Figure~\ref{fig:absorbHb}), with a larger uncertainty due to its degeneracy with the absorption. This quantity will be used for metallicity estimations.}
\end{deluxetable}

%%%%%%%% Figure 4 %%%%%%%%%%%%%%%%%%%%%%%%%%%%%%%%%%%%%%%%%%%

\subsection{Balmer Decrement}
\label{sec:decrement}

The broad Balmer decrements of LRDs are high: the median broad
H$\alpha$/H$\beta$ is 7.6 at $2\lesssim z\lesssim6$ \citep{stone2026}. Interpreted 
as caused by dust reddening, these decrements imply up to several magnitudes of visual extinction toward the
broad-line region, and a correction of the same size to the luminosities of
the broad lines. Dense, nearly dust-free gas (Section~\ref{sec:intro}) 
can also produce a high intrinsic decrement, because the Case~B value 
of 2.86 \citep{osterbrock2006} does not apply: the Balmer lines become optically thick and collisional excitation  
can raise
the decrement \citep{torralba2026}. In the shadowing model the intrinsic
broad H$\alpha$/H$\beta$ is 4.6, and 10.6 after the dust of the equatorial
clouds \citep{madau2026a}. The decrement of QSO1 therefore depends on whether its
broad lines are reddened by dust or emitted by dense gas.
Only in the former case can they be used to infer an extinction correction.

 We computed the line ratios of dust-free gas illuminated by an AGN with \textsc{Cloudy} \citep{gunasekera2023}, over $n_{\rm H}=10^{2}$--$10^{12}$\,cm$^{-3}$ (Appendix~\ref{append:cloudy}; Figure~\ref{fig:balmer}). Below $n_{\rm H}\simeq10^{4}$\,cm$^{-3}$ the ratios are those of Case~B; between $10^{5}$ and $10^{8}$\,cm$^{-3}$ H$\gamma$/H$\alpha$ and H$\delta$/H$\alpha$ lie above the values for dust, and the higher Balmer lines separate dense gas from reddening; above $\simeq10^{8.5}$\,cm$^{-3}$ the ratios follow the reddening curve and dust and dense gas cannot be independently determined.

Images B and C add a test in time. Dust at the radius of the broad-line
region, a few tens of light-days from the black hole \citep{bentz2013}, moves by
$\simeq9$ light-days in 2.4\,yr at $3000\kms$, so the dust column toward the
broad-line region can change between the images only if it is patchy on this
scale. The density and the illumination of the gas can change on shorter
timescales. A decrement that differs between the images more likely points to the gas, or, for the broad lines, to differential microlensing of the broad-line region (Section~\ref{sec:microlensing}).\par

We measure H$\alpha$, H$\gamma$ and H$\delta$ relative to H$\beta$ in
images B and C, for both the broad and the narrow components. The
narrow lines come from gas of low density, where Case~B applies, so a
narrow-line decrement close to 2.86 would place any reddening in the
broad-line region. Narrow H$\alpha$ is not detected in image C, and the narrow
H$\alpha$/H$\beta=2.6_{-0.6}^{+0.8}$ in image B (Table~\ref{tab:other}) is
consistent with Case~B and with the narrow-line $A_V$ of
\citet{deugenio2025}. With the intermediate component treated
separately (Section~\ref{sec:broad}), the broad H$\alpha$/H$\beta$ is
$5.1\pm0.3$ and $3.2\pm0.7$ in B and C, which corresponds to
$A_V=1.5\pm0.2$\,mag for the SMC extinction curve. Including the
intermediate component gives $9.4_{-0.4}^{+0.5}$ and
$6.3_{-0.8}^{+0.9}$ (weighted mean $8.7\pm0.4$, $A_V=3.1\pm0.1$\,mag;
Table~\ref{tab:lines}), and $10.6^{+1.7}_{-1.4}$ in the SPURS
spectrum, against the published $7.0\pm2.3$ \citep{tang2026}; this
$A_V$ is consistent with the independent result of
\citet{deugenio2025}, $A_V=2.9\pm0.2$\,mag. 
However, how the decrement is divided between the intermediate component and the rest of the broad line is model-dependent. There are Balmer absorptions in the core. At $R\simeq1000$ the absorption and an H$\beta$ counterpart of the intermediate component are degenerate, and the data cannot constrain both simultaneously. The counterpart is absent and the intermediate component has H$\alpha$/H$\beta>28$ ($3\sigma$) only if no strong absorption is included. At $R\simeq2700$, \citet{deugenio2025} fit broad H$\beta$ and H$\alpha$ with the same two-Gaussian profile and a partially covering absorber, and obtain H$\alpha$/H$\beta=9.1\pm0.8$, although an H$\beta$ absorber much stronger than H$\alpha$ is needed. The two limiting cases lead to different conclusions. If the intermediate component has no counterpart in H$\beta$, it carries a high decrement and the rest of the broad line requires at most $A_V\simeq1.5$\,mag. If the counterpart is hidden by the absorption, the decrement is $\approx9$ at all velocities and the broad-line region requires either $A_V\simeq3$\,mag or gas denser than $10^{8.5}\,{\rm cm^{-3}}$. Although the profiles alone do not distinguish the two,
the first case is favored: hiding the counterpart would
require an H$\beta$ absorber substantially stronger
than the one we fit, and a stronger absorption still in
H$\alpha$, since both lines absorb from the $n=2$
level, while the observed H$\alpha$ absorption is too
weak. This motivates the fiducial decomposition of
Section~\ref{sec:broad}; we quote both cases where they
differ.
 
Including the 15\% slit-loss systematic
(Section~\ref{sec:data}), the decrements of B and C differ by
$1.6\sigma$, with or without the intermediate component. 
H$\gamma$/H$\alpha$ and H$\delta$/H$\alpha$ are $<0.02$
and $0.012\pm0.003$, respectively. The broad H$\delta$ is tentative because its detection significance is sensitive to the fitting window; a shorter window would elevate the continuum and lower the significance below $3\sigma$. Nevertheless, its upper limit remains tight (H$\delta$/H$\alpha<0.02$). These constraints exclude
dust-free gas at $10^{5}$--$10^{8}$\,cm$^{-3}$, for which
H$\gamma$/H$\alpha$ and H$\delta$/H$\alpha$ would lie above the reddening curve
(Figure~\ref{fig:balmer}), so the broad-line region requires either reddening of $A_V\simeq1.5$--$3$\,mag, depending on the decomposition, or gas denser than $10^{8.5}\,{\rm cm^{-3}}$; the higher Balmer lines alone do not distinguish the
two. Most of the broad-line ratios of other LRDs lie on or close to
the same curve \citep{tang2026b, nikopoulos2026b}. The decrement alone therefore does not decide between dust and dense 
gas. The He\,\textsc{i} excess and the Balmer absorption, which do 
not depend on the decomposition, point to dense gas 
(Section~\ref{sec:ionizing}), and at the low metallicity of QSO1 
(Section~\ref{sec:metal}) little dust is expected to form 
(Section~\ref{sec:pristine}). 

%%%%%%%% Figure 5 %%%%%%%%%%%%%%%%%%%%%%%%%%%%%%%%%%%%%%%%%%%

\subsection{The Narrow-line Region and the Metallicity of the Gas}
\label{sec:metal}
\label{sec:other}

The narrow lines come from gas far from the black hole. They test whether
an AGN ionizes that gas and measure its metallicity.
Table~\ref{tab:other} lists their fluxes or $3\sigma$ limits in each image.

Narrow He\,\textsc{ii}/H$\beta$ measures whether photons above 54\,eV reach
the narrow-line region. We find a loose limit of He\,\textsc{ii}/H$\beta<0.6$. Values of $\approx0.1$ and above are found in
narrow-line regions ionized by an AGN \citep{ubler2023}. In our photoionization models (Appendix~\ref{append:cloudy}), an AGN with weak X-ray emission gives 0.01--0.02, so values down to 0.01 are compatible with an AGN.
[Ne\,\textsc{v}] $\lambda3426$ and [Fe\,\textsc{vii}] $\lambda6087$ require
photons above 97 and 99\,eV, which stars produce in negligible numbers, and
their detection would identify a narrow-line region ionized by the AGN.
Neither is detected, with loose limits of $<0.5$ relative to narrow H$\beta$.

The ratio [O\,\textsc{iii}] $\lambda5007/\lambda4363$ depends on the electron
temperature and, at high density, on the density. At the low metallicity of
the host \citep{maiolino2025pristine} the line ratio is $\approx20$--40 at densities
$\lesssim10^{4}$\,cm$^{-3}$, where it is primarily set by the temperature alone; smaller values require densities above the critical
density of $\lambda5007$, as found in luminous LRDs \citep{wang2026b}. 
We obtain a loose limit of $\lambda5007/\lambda4363 > 1.5$, consistent with
gas below $10^{5}$\,cm$^{-3}$.

The metallicity of the gas around QSO1 bears on how its black hole
formed. \citet{maiolino2025pristine} measured
$R_3=$\,[O\,\textsc{iii}]$\lambda5007$/H$\beta\simeq0.55$ in image A with the
NIRSpec IFU at $R\simeq2700$, which gives $12+\log({\rm O/H})\simeq6.4$, or
$Z\simeq5\times10^{-3}\,Z_\odot$, with the calibration of \citet{sanders2024}.
This is among the lowest metallicities measured in an LRD; the typical value is
$\simeq0.08\,Z_\odot$ \citep{nikopoulos2026}. One caveat is that strong-line calibrations such as that of
\citet{sanders2024} are derived from star-forming galaxies and are
extrapolated at these metallicities; their applicability to QSO1 is
examined in Appendix~\ref{append:cloudy}.\footnote{Unfortunately, [N\,\textsc{ii}]\,$\lambda$6583 is blended with broad H$\alpha$ line and
cannot be reliably measured. Furthermore, at $z=7.04$ [S\,\textsc{ii}]\,$\lambda\lambda$6716,
6731 falls at 5.40--5.41\,$\mu$m, near the red end of G395M and it is
undetected.}

%%%%%%%% Figure 6 %%%%%%%%%%%%%%%%%%%%%%%%%%%%%%%%%%%%%%%%%%%
\begin{figure}
\centering
\includegraphics[width=\columnwidth]{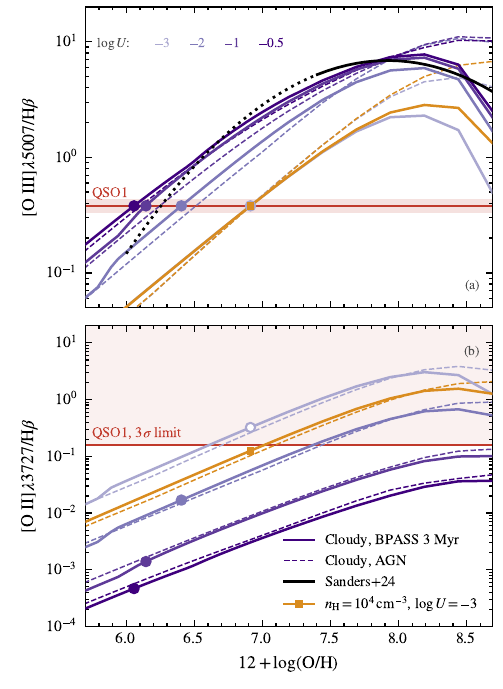}
\caption{Oxygen abundance from the narrow lines. (a)
[O\,\textsc{iii}]$\lambda5007$/H$\beta$ against $12+\log({\rm O/H})$ in our
\textsc{Cloudy} models at $n_{\rm H}=10^{2}$\,cm$^{-3}$, ionized by a 3\,Myr
BPASS population (solid) and by the AGN continuum (dashed), for four values
of $\log U$ (shades).
The 10\,Myr population and the AGN with weak X-ray emission
(Section~\ref{sec:metal}) are not shown; at $R_3=0.38_{-0.05}^{+0.06}$ they give O/H within
0.1\,dex of the 3\,Myr and AGN models, and up to 0.25\,dex higher at
$\log U=-3$.
The black curve is the calibration of
\citet{sanders2024}, dotted where it is extrapolated. The red band is the
$R_3$ of QSO1, and the circles mark where each BPASS model matches it.
(b) [O\,\textsc{ii}]$\lambda3727$/H$\beta$ for the same models and the same
circles. The shaded region lies above the $3\sigma$ limit of QSO1; open
circles fall in it and are excluded. The orange curves and squares show the same models at $n_{\rm H}=10^{4}\,{\rm cm^{-3}}$ for $\log U=-3$: in panel (a) they overlap with the $n_{\rm H}=10^{2}\,{\rm cm^{-3}}$ curves at low metallicity, while in panel (b) they fall below the limit, so that $\log U=-3$ is allowed at this density.
}
\label{fig:metal}
\end{figure}

We compare the measured ratios with a grid of \textsc{Cloudy} models of ionization-bounded, dust-free gas, with four ionizing spectra and the metallicity, the ionization parameter and the density free (Appendix~\ref{append:cloudy}; Figure~\ref{fig:metal}).
QSO1 shows narrow Balmer absorption \citep{ji2025,
maiolino2025pristine, deugenio2025}, which must be accounted for in
the narrow-line ratios. At $R\simeq1000$ the narrow emission and the
absorption are difficult to separate, and only image B shows the
absorption clearly (Figure~\ref{fig:absorbHb}). We therefore measure
$R_3$ in image B only, with the absorption component included (unlike
in Table~\ref{tab:lines}), and obtain $R_3=0.38_{-0.05}^{+0.06}$,
consistent with the value of $0.55\pm0.16$ reported by
\citet{maiolino2025pristine}.

The fit gives [O\,\textsc{ii}]$\lambda3727$/H$\beta<0.16$,
[Ne\,\textsc{iii}]$\lambda3869$/H$\beta<0.15$, and narrow
He\,\textsc{ii}\,$\lambda4686$/H$\beta<0.24$. The [O\,\textsc{ii}]
limit sets a lower bound on $U$ that depends on the density:
$\log U\gtrsim-2.8$ at $n_{\rm H}=10^{2}$\,cm$^{-3}$ and $\gtrsim-3.1$
at $10^{4}$\,cm$^{-3}$, where collisional de-excitation weakens
[O\,\textsc{ii}] (Figure~\ref{fig:metal}b). Within these bounds O/H is
set by $R_3$. The limits on [O\,\textsc{iii}]$\lambda4363$,
O\,\textsc{iii}]$\lambda1666$ and narrow He\,\textsc{ii}\,$\lambda4686$
are too loose to constrain the density or the ionizing source. The
models give $12+\log({\rm O/H})=6.0$--7.1, or
$Z=2$--$26\times10^{-3}\,Z_\odot$, for all the ionizing spectra
allowed. The upper end is reached at $10^{4}$\,cm$^{-3}$; at higher
densities the [O\,\textsc{ii}] limit no longer constrains $U$, and O/H
could be higher (Section~\ref{sec:pristine}). For a stellar mass
$M_\star<2\times10^{7}$\,\msun\ \citep{juodzbalis2026}, the range 6.0--7.1  lies up
to 0.3--1.4\,dex below the mass--metallicity relation at $z\sim6$--8
\citep{chemerynska2024}; QSO1 falls on the relation only if
$M_\star\lesssim4\times10^{6}$\,\msun.
\begin{figure*}
\centering
\includegraphics[width=0.8\textwidth]{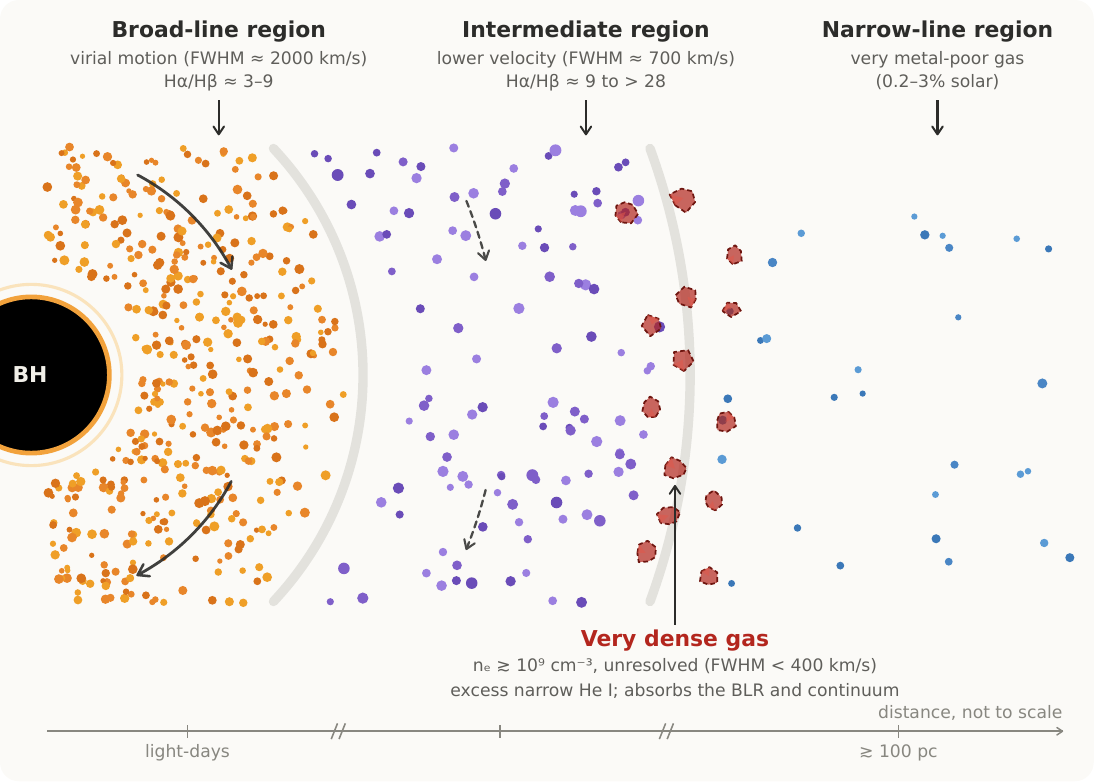}
\caption{Schematic of the environment of the black hole in QSO1 as
constrained by this work (not to scale). Orange: the broad-line region,
a few tens of light-days from the black hole, whose broad H$\beta$ and
H$\alpha$ widths trace virial motion (Section~\ref{sec:widths}). Its
Balmer decrement is H$\alpha$/H$\beta\approx3$--5 with the intermediate
component of H$\alpha$ fitted separately, and $\approx9$ with it
included (Section~\ref{sec:decrement}). Purple: the intermediate region,
which emits the H$\alpha$ component of FWHM\,$\simeq700$\,\kms\ with no
counterpart of the same width in H$\beta$ (Section~\ref{sec:broad}). Its
decrement ranges from $\approx9$, if the H$\beta$ counterpart is hidden
by the narrow absorption \citep{deugenio2025}, to $>28$ if it is absent;
it is drawn outside the broad-line region, as its lower velocity
implies for virial motion. Dashed arrows indicate its velocity scale,
not a measured rotation. Red: very dense gas ($n_{\rm e}\gtrsim10^{9}$
\,cm$^{-3}$, unresolved, FWHM\,$<400$\,\kms), traced by the excess of
narrow He\,\textsc{i}\,$\lambda5876$ and by the narrow Balmer
absorption, which covers the broad-line region and the continuum
(Sections~\ref{sec:fitting} and \ref{sec:ionizing}); its location is
otherwise unconstrained. We speculate that the intermediate region and the very
dense gas are the precursors of the torus of ordinary
quasars (Section~\ref{sec:pristine}). Blue: the very metal-poor narrow-line region,
$12+\log({\rm O/H})=6.0$--7.1, or 0.2--3\% solar
(Section~\ref{sec:metal}), on scales $\gtrsim100$\,pc
(Appendix~\ref{append:lensing}).}
\label{fig:sketch}
\end{figure*}
\section{Discussion}
\label{sec:discussion}

We now ask what the measurements of Section~\ref{sec:analysis}
reveal about the gas around the black hole in QSO1: how it moves
(Section~\ref{sec:widths}), whether it is dense or reddened by dust
(Section~\ref{sec:ionizing}), and how enriched it is
(Section~\ref{sec:pristine}). Figure~\ref{fig:sketch} summarizes the
structure that emerges: a broad-line region in virial motion, an
intermediate region at lower velocity, very dense low-dispersion gas,
and a very metal-poor narrow-line region.

\subsection{The Origin of the Broad-line Widths}
\label{sec:widths}

Three explanations of the line widths of LRDs predict different ratios
of the H$\beta$ to H$\alpha$ widths. Electron scattering in a single
layer of gas gives every line the same width \citep{rusakov2026}. In a
broad-line region stratified in radius, H$\beta$ is emitted closer to
the black hole than H$\alpha$ and is broader by about 20\%
\citep{bentz2010, madau2026wings}. In a dense cocoon, bound-free
opacity shortward of the Paschen edge removes the wings of H$\alpha$,
and H$\beta$ is broader by 50--60\% \citep{sneppen2026}.

With the intermediate component fitted separately
(Section~\ref{sec:broad}), the line dispersions of broad H$\beta$ and
broad H$\alpha$ are equal within the uncertainties, consistent both
with a stratified broad-line region and with a single scattering
layer. The widths alone therefore do not exclude electron scattering,
but the black hole mass provides a discriminating test. In the
scattering model the observed width is set mostly by electron
scattering and the underlying orbital velocities are much lower; since
the virial mass scales as the square of the velocity, the inferred
mass is about 2\,dex lower than the virial estimate from the same
lines \citep{rusakov2026, deugenio2025}. The rotation curve of the
central region in QSO1 gives a dynamical mass of
$10^{7.7\pm0.3}$\,\msun\ \citep{juodzbalis2026}; this value is
consistent with our virial masses from broad H$\beta$ and broad
H$\alpha$ (Section~\ref{sec:broad}) within the scatter of single-epoch
estimators. We therefore conclude that, once the intermediate
component is excluded, the widths of broad H$\beta$ and broad
H$\alpha$ trace virial motion around the black hole.

\subsection{Dense Gas or Dust}
\label{sec:ionizing}

The data provide three independent signatures of very dense gas. The
narrow He\,\textsc{i}~$\lambda5876$/H$\beta$ is $1.4_{-0.2}^{+0.3}$, or $0.5\pm0.1$ with the absorption included in H$\beta$
(Section~\ref{sec:heii}), several times the recombination value of
$\simeq0.1$ \citep{porter2012}. Reddening cannot produce the excess:
the narrow lines show no evidence of dust
(Section~\ref{sec:decrement}), and even $A_V\simeq3$\,mag with the SMC
extinction curve would raise the ratio by only a factor of two. The
excess requires collisional excitation from the metastable
$2\,^3$S level of He$^0$, which enhances $\lambda5876$ at
$n_e\gtrsim10^{9}$\,cm$^{-3}$ \citep{benjamin1999, porter2012}; the
same level absorbs $\lambda3889$, consistent with
$\lambda3889/\lambda5876<0.5$, below the recombination value of
$\simeq0.8$. The narrow Balmer absorption in H$\beta$
(Section~\ref{sec:fitting}; Figure~\ref{fig:absorbHb}) likewise requires a large
population of hydrogen in the $n=2$ level, which is maintained only
in dense, optically thick gas \citep{inayoshi2025b, ji2025}.
The permitted Fe\,\textsc{ii}\,$\lambda1786$ multiplet, reported by \citet{tang2026} in the SPURS spectrum of image B, is also detected in the independent DAJE spectrum of the same image (Table~\ref{tab:other}). This multiplet is likely excited by Ly$\alpha$ fluorescence in gas that is optically thick to Ly$\alpha$ \citep{sigut2003,tang2026}, consistent with the dense gas inferred from the Balmer absorption; like He\,\textsc{i}, the line is narrow and therefore arises outside the broad-line region.

The kinematics constrain where this gas lies. The He\,\textsc{i} line
is unresolved ($\sigma\lesssim170$\,km\,s$^{-1}$), whereas gas in
virial motion at the radius of the broad-line region moves at
thousands of km\,s$^{-1}$; and the absorber covers the broad H$\beta$
emission as well as the continuum (Section~\ref{sec:fitting}). The
dense gas therefore lies outside the broad-line region (Figure~\ref{fig:sketch}), either at
larger radii than the intermediate component or in a structure that is
not in virial motion, such as a pressure-supported envelope or a slow
inflow or outflow. The data do not distinguish between the two.

The Balmer decrement is consistent with dense gas but does not require it. If the intermediate component of H$\alpha$ has no counterpart in H$\beta$, most of the decrement arises in the intermediate component, with
H$\alpha$/H$\beta>28$, while the broad-line region requires at most
$A_V\simeq1.5$\,mag (Section~\ref{sec:decrement}). Interpreted as dust, the
intermediate component would need $A_V\gtrsim6$\,mag for the SMC
curve, distributed so as to leave the broad-line region behind it
nearly unreddened. Optically thick gas produces such decrements
without dust: in our \textsc{Cloudy} models, H$\alpha$/H$\beta$
reaches $\simeq100$ in light emerging through the shielded face at
$n_{\rm H}\simeq10^{7}$--$10^{8}$\,cm$^{-3}$
(Appendix~\ref{append:cloudy}). If instead the counterpart is hidden by the Balmer absorption, the decrement is $\approx9$ throughout the broad line, which dust with $A_V\simeq3$\,mag and gas denser than $10^{8.5}\,{\rm cm^{-3}}$ explain equally well. The evidence for dense gas therefore rests on the He\,\textsc{i} excess and on the Balmer absorption, which hold in either case.

The helium lines could in principle also test the ionizing spectrum
that reaches the gas. If a geometrically thick, super-Eddington disk
shadows the gas from its hardest photons \citep{madau2026a,
madau2026b}, He\,\textsc{ii} is weak while He\,\textsc{i} keeps its
usual strength; if a layer of dense gas filters the spectrum
\citep{inayoshi2025b, ji2025, zucchi2026}, both lines are suppressed.
For QSO1, the shadowing models predict broad
He\,\textsc{ii}~$\lambda4686$/H$\beta$ of 0.20--0.27 at 12 times the
critical accretion rate and 0.08--0.13 at 32 times it
\citep{madau2026a}. Our limit, $<0.14$, lies below the first range
and above the second. The broad He\,\textsc{i}/H$\beta<0.13$, or
$\lesssim0.09$ corrected for $A_V=1.5$\,mag, is comparable to the
value of $\simeq0.1$ in type~1 AGN \citep{vandenberk2001}. The
current limits therefore do not yet distinguish shadowing from
filtering.

\subsection{Metallicity, Host, and the Missing Torus}
\label{sec:pristine}

The narrow-line gas of QSO1 is metal-poor. Our models give
$12+\log({\rm O/H})=6.0$--7.1, or $Z=2$--$26\times10^{-3}\,Z_\odot$
(Section~\ref{sec:metal}), in agreement with the value of
$Z\simeq5\times10^{-3}\,Z_\odot$ measured by
\citet{maiolino2025pristine} and below the typical value of
$\simeq0.08\,Z_\odot$ for LRDs \citep{nikopoulos2026}. The models assume
gas at $n_{\rm H}\le10^{4}$\,cm$^{-3}$; if the [O\,\textsc{iii}]
emission arose at $\sim10^{6}$\,cm$^{-3}$, the same $R_3$ would require
0.3--0.7\,dex more oxygen (Appendix~\ref{append:cloudy}).

The host is also small relative to the black hole: its stellar mass 
($<2\times10^{7}$\,\msun; \citealt{juodzbalis2026}) is below the dynamical black hole mass of
$\simeq5\times10^{7}$\,\msun\ and the gas lies
below the mass--metallicity relation at $z\sim6$--8
(Section~\ref{sec:metal}). Few stars have therefore formed and
enriched the gas around the black hole, which points to a black hole
that grew faster than its host. Two classes of seeding models are consistent with this. A direct-collapse 
black hole growing inside a nearly pristine inflow reaches
$Z\simeq5\times10^{-3}\,Z_\odot$ after 75\,Myr \citep{pacucci2026},
though at $\simeq3\times10^{6}$\,\msun, an order of magnitude below
QSO1; heavy seeds instead form mostly in gas at
$Z\sim10^{-3}$--$10^{-2}\,Z_\odot$ \citep{chon2025, chon2026b,
prole2026}. Our range overlaps both, so the metallicity alone cannot
distinguish them.

The low metallicity may also explain why LRDs lack the hot dust of
ordinary quasars \citep{setton2025}. In ordinary AGN the dust
sublimation radius separates the dust-free broad-line region from the
dusty torus, which may be the outer part of the same flow
\citep{elitzur2006}. At 0.2--3\% solar little dust is expected, and the
gas outside the broad-line region can remain dust-free. We therefore
speculate that the intermediate region and the very dense gas of QSO1
are the precursors of the torus: gas at the same location, at velocities
that decrease outward if their motion is virial, before dust has
formed. As star formation enriches the gas, radiation pressure on the
newly formed dust, a few hundred times stronger than on free electrons
\citep{fabian2008}, would remove the gas of low column density and
leave a thick, dusty torus. If this picture is correct, Balmer
absorption and the narrow He\,\textsc{i} excess should become rarer,
and hot dust emission more common, among more metal-rich LRDs.

\section{Summary}
\label{sec:summary}
We have presented deep NIRSpec spectroscopy at $R\simeq1000$ of images B
and C of the lensed little red dot A2744-QSO1 from the DAJE program, with
exposures up to 13 times those of the deepest previous medium-resolution spectrum, together
with archival JWST spectra reduced and fitted in the same way. The DAJE
spectrum of image A is contaminated and will be analyzed in future work.
Our main results are:

\begin{itemize}

\item Image A is 0.3--0.7\,mag brighter than all lens models predict, in
the rest-frame ultraviolet and optical alike, while B and C agree with
them. Stellar microlensing cannot produce this excess, while millilensing by
dark-matter halos and subhalos can account for the excess.

\item Over the 2.8 rest-frame years sampled by seven spectra, the
H$\beta$ and H$\alpha$ equivalent widths show no significant
variability ($\sigma_0/m<32\%$ and $<15\%$; 95\% upper limits). The changes reported in the literature are consistent with the
measurement uncertainties. Within the errors, the equivalent widths would not
detect the variability of an ordinary quasar. The H$\beta$
equivalent width of image C is $52\pm16\%$ higher than in the earlier
spectra; with the current data, a change of the source cannot be
separated from a systematic specific to image C.

\item H$\alpha$ contains an intermediate component with
${\rm FWHM}\simeq700$\,km\,s$^{-1}$ and no detected
counterpart in H$\beta$, which likely arises outside
the broad-line region. Once it is excluded, broad
H$\beta$ and broad H$\alpha$ have comparable widths (${\rm FWHM}\simeq2000$\,km\,s$^{-1}$) and
yield virial masses consistent with the dynamical black
hole mass of $10^{7.7}$\,\msun. We conclude that they
are broadened by virial motion around the black hole.

\item We detect He\,\textsc{i}~$\lambda5876$ in
QSO1. The line is narrow, and its ratio to narrow H$\beta$ is several
times the recombination value; together with the narrow Balmer
absorption, it requires very dense gas
($n_e\gtrsim10^{9}$\,cm$^{-3}$) with a velocity dispersion
$\lesssim170$\,km\,s$^{-1}$, located outside the broad-line region. We also confirm the narrow Fe\,\textsc{ii}\,$\lambda1786$ emission reported previously in SPURS data, which is likely excited by Ly$\alpha$ fluorescence in gas optically thick to Ly$\alpha$, consistent with the dense-gas scenario.

\item The broad H$\alpha$/H$\beta$ ratio is 9.4 and 6.3 in images B and C. How this decrement is divided between the intermediate component and the broad-line region depends on the model, because the core of H$\beta$ is absorbed. If the intermediate component has no counterpart in H$\beta$, it carries most of the decrement (H$\alpha$/H$\beta>28$) and the broad-line region requires at most $A_V\simeq1.5$\,mag; if the counterpart is hidden by the absorption, the decrement is $\approx9$ throughout the broad line, which requires $A_V\simeq3$\,mag or gas denser than $10^{8.5}$\,cm$^{-3}$. The decrement alone does not distinguish dense gas from dust. The current limits on He\,\textsc{ii} and broad He\,\textsc{i}
do not yet distinguish whether the ionizing spectrum is shadowed by a
thick disk or filtered by dense gas.

\item Photoionization models of the narrow lines give
$12+\log({\rm O/H})=6.0$--7.1. The gas is metal-poor, and the host
stellar mass is below the black hole mass. This suggests that the black hole has grown in
a chemically young environment, consistent both with growth inside a
nearly pristine inflow and with heavy seeds formed in weakly enriched
gas. At this metallicity little dust is expected, and we
speculate that the intermediate region and the very
dense gas (Figure~\ref{fig:sketch}) are the precursors
of the torus of ordinary quasars, before dust has
formed.

\end{itemize}
\begin{acknowledgments}
This work is based on observations made with the NASA/ESA/CSA James
Webb Space Telescope. The data were obtained from the Mikulski
Archive for Space Telescopes at the Space Telescope Science
Institute, which is operated by the Association of Universities for
Research in Astronomy, Inc., under NASA contract NAS5-03127 for
JWST. These observations are associated with program \#9645. Support
for US investigators was provided by NASA through a grant from the
Space Telescope Science Institute, which is operated by the
Association of Universities for Research in Astronomy, Inc., under
NASA contract NAS5-03127.
This work was carried out within the COSMOWEBB project ``A new era
for cosmology: exploiting the JWST revolution'', funded by the
Italian National Institute for Astrophysics (INAF) through the PRORIS
programme (PROgramma di RIcerca Spaziale di base), CUP C83C26000300005.
RSR acknowledges support from the Science and Technology Facilities
Council (STFC) through a PhD studentship under grant number UKRI1789.
DG acknowledges support from the Kavli Foundation under Grant
No. PS-2025-GR-249-3072, and thanks the Brinson Foundation for their
continued support of his research.
PSh acknowledges support from DoE Grant DE-SC0026113.
LB acknowledges the support by the Italian Ministry of University and
Research (grant FIS2023-01611, CUP C53C25000300001).
PB acknowledges financial support from the Italian Space Agency (ASI)
through contract ``Euclid - Phase E'', and from INAF through the
AF2025 grants ``PARSE'' and ``OpenLens-Vet'', and the RF2024 grant
``Probing Dark Matter and Galaxy Formation in Galaxy Clusters through
Strong Gravitational Lensing''.
MB acknowledges support from the ERC Grant FIRSTLIGHT \# 101053208,
Slovenian national research agency ARIS through grants N1-0238 and
P1-0188, and ESA PRODEX Experiment Arrangements No. 4000146646 and
4000149972.
MC acknowledges support from the INAF GO Grant 2024 ``Revealing the
nature of bright galaxies at cosmic dawn with deep JWST
spectroscopy''.
SC acknowledges support from the MUR projects FIS-2024-01621 DAWN.
KG acknowledges support from Australian Research Council Laureate
Fellowship FL180100060.
MMen acknowledges financial support from INAF through grants AF2025
``PARSE'' and ``OpenLens-Vet'' and RF2024 ``Probing Dark Matter and
Galaxy Formation in Galaxy Clusters through Strong Gravitational
Lensing''.
EV and MMes acknowledge support from INAF -- Astrofisica di Frontiera
2025, Excellence Track grant ``PARSE'' (PI Vanzella).
TM acknowledges support by JSPS KAKENHI Grant Number 25K24918.
BPD and PSa acknowledge financial support from INAF RF2024 Large
Grant ``UNDUST: UNveiling the Dawn of the Universe with JWST''.
RS acknowledges support from the MUR projects FIS-2024-01621 DAWN and
the INFN TEONGRAV initiative.
MT acknowledges support from the ARC Discovery Project DP260100569.
Anthropic's Claude was used for copyediting, the setup of the
\textsc{Cloudy} grids and the plotting scripts; all results were
checked and the scientific analysis and conclusions are the authors'
own.
This work made use of v2.2.1 of the Binary Population and Spectral
Synthesis (BPASS) models as described in \citet{eldridge2017} and
\citet{stanway2018}.
\end{acknowledgments}

\facilities{JWST (NIRSpec, NIRCam)}

\software{\texttt{jwst} \citep{Bushouse2025},
\texttt{msaexp} \citep{brammer2023},
\texttt{msafit} \citep{degraaff2024msafit},
\texttt{PySersic} \citep{pasha2023},
\texttt{unite} \citep{hviding_unite},
\textsc{Cloudy} C23.01 \citep{gunasekera2023},
Astropy \citep{astropy2013, astropy2018, astropy2022},
NumPy \citep{harris2020},
Matplotlib \citep{hunter2007},
\texttt{dust\_extinction} \citep{gordon2024}}

\appendix
\section{Comparison with published measurements}
\label{append:comparison}
We reanalyze the archival UNCOVER PRISM and SPURS grating spectra, and in this appendix we compare the measurements with the published ones for H$\alpha$ and H$\beta$, the two Balmer lines discussed most extensively in previous works. The absolute flux scale of an MSA spectrum depends on the slit-loss correction and on the adopted normalization, and differs between reductions by tens of percent; we therefore compare rest-frame EWs and line ratios, which are independent of it.\par
\textit{UNCOVER}. The EWs agree with those of \citet{furtak2025} for all four PRISM spectra, with a different line-profile model. For H$\alpha$ we measure $286^{+6}_{-5}$, $289^{+8}_{-8}$, $313^{+7}_{-7}$ and $381^{+22}_{-20}$\,\AA\ for A1, A2, B and C, against $285^{+7}_{-6}$, $289^{+9}_{-8}$, $307^{+4}_{-5}$ and $346^{+11}_{-11}$\,\AA; for H$\beta$, $45^{+4}_{-4}$, $56^{+7}_{-6}$, $46^{+3}_{-3}$ and $72^{+11}_{-10}$\,\AA, against $48^{+2}_{-2}$, $53^{+4}_{-4}$, $50^{+2}_{-2}$ and $62^{+5}_{-5}$\,\AA. All these values agree within $2\sigma$. The two Gaussians of their H$\alpha$ model for A1, with FWHM $=800\pm60$ and $2533\pm120$\,km\,s$^{-1}$, closely match the narrow and the wide Gaussians of $H_{\alpha}$ in Table~\ref{tab:lines}.\par
\textit{SPURS}. \citet{tang2026} fitted each Balmer line with one narrow and one broad Gaussian. The EWs of [O\,\textsc{iii}]\,$\lambda5007$ ($4.8\pm1.4$\,\AA) and of H$\beta$ ($41^{+6}_{-6}$\,\AA) agree with theirs ($5.0\pm1.5$ and $35\pm7$\,\AA) within $1\sigma$. The H$\alpha$ EW, $342^{+24}_{-22}$\,\AA, is 1.7 times the $204\pm48$\,\AA\ (broad plus narrow) of their Table~B2. Their plotted spectrum, digitized, gives an H$\alpha$ EW of $\approx300$--400\,\AA, consistent with the value measured here. The line model may partly explain the lower tabulated EW: fitting the spectrum with one narrow and one broad Gaussian gives a total H$\alpha$ EW of $298^{+19}_{-17}$\,\AA, with a worse fit. The broad H$\alpha$/H$\beta$ is correspondingly higher, $10.6^{+1.7}_{-1.4}$ against their $7.0\pm2.3$, and the two agree within $2\sigma$.\par
\textit{BlackTHUNDER}. This IFU spectrum of image~A \citep{ji2025, deugenio2025} was taken in a different NIRSpec observing mode from the multi-object spectroscopy (MOS) used here. To avoid systematic differences between the two modes, we use only MOS data in the main text and compare with the published IFU measurements here. The two analyses of the IFU data do not agree on the broad H$\beta$ EW: \citet{ji2025} reported $72.6^{+10.1}_{-7.7}$\,\AA\ and \citet{deugenio2025} $38^{+6}_{-5}$\,\AA. The values in Table~\ref{tab:lines}, $41$--$72$\,\AA, lie between them. \citet{deugenio2025} also measured a broad H$\alpha$ EW of $170^{+20}_{-10}$\,\AA, smaller than the $286$--$381$\,\AA\ of Table~\ref{tab:lines}. The line fluxes differ much less. Their broad H$\alpha$ flux, $1.39^{+0.09}_{-0.06}\times10^{-17}$\,erg\,s$^{-1}$\,cm$^{-2}$, is consistent with that of image~A in the first UNCOVER epoch ($1.40^{+0.02}_{-0.02}$ in the same units), and the broad H$\beta$ fluxes agree within a factor of 1.5. The EW differences come primarily from the continuum. For this reason, we do not include the BlackTHUNDER measurements in the variability analysis. \par
The H$\beta$ absorption in the DAJE spectrum of image B is modeled with a
Gaussian optical depth
$\tau(v)=\tau_0\exp[-(v-\Delta v_{\rm abs})^2/2\sigma_{\rm abs}^2]$, where
$\tau_0$ is the optical depth at the center of the trough, $\Delta v_{\rm
abs}$ its velocity offset and $\sigma_{\rm abs}$ its velocity dispersion;
the broad component and the continuum are multiplied by $e^{-\tau}$. The
fitted properties are $\Delta v_{\rm abs}=27_{-14}^{+15}$\kms,
$\sigma_{\rm abs}<141$\kms\ and EW\,$=9.8_{-1.2}^{+1.3}$\,\AA, against
$\Delta v_{\rm abs}=50_{-20}^{+20}$\kms, $\sigma_{\rm
abs}=110_{-10}^{+20}$\kms\ and EW\,$=11_{-2}^{+3}$\,\AA\ in
\citet{deugenio2025}. No Balmer absorption is apparent in the SPURS and
DAJE C spectra or in H$\alpha$ of DAJE B, where BlackTHUNDER reported an
absorber with a smaller EW. When the absorber is included, the broad
H$\beta$ flux increases by only $\sim10\%$, while the narrow flux
increases by a factor of several. If the absorber lies in the dense gas
close to the black hole, its depth can change between the epochs sampled
by the images; empirically, the parameters measured here agree with those
from BlackTHUNDER at a different epoch, and, since the absorption changes
the broad H$\beta$ flux by only $\sim10\%$, a change of the absorber would
enter the comparison of the images at the level of a few percent.

\begin{figure}
\centering
\includegraphics[width=\linewidth]{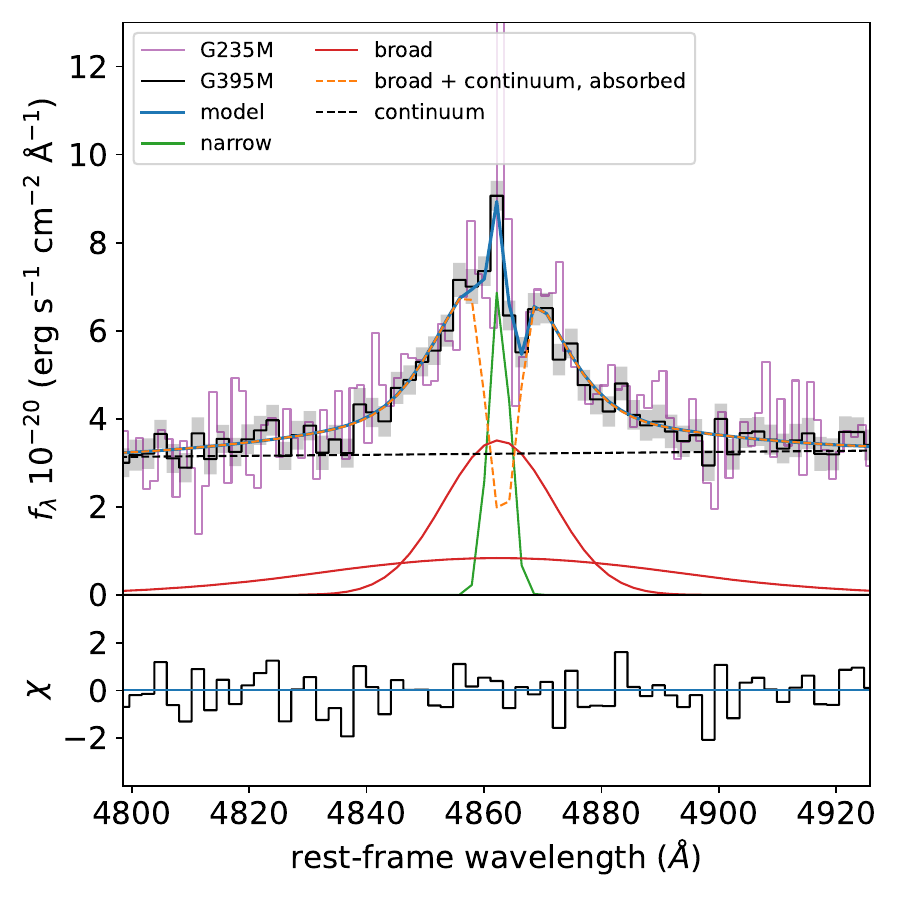}
\caption{Best-fit H$\beta$ profile with an absorption component included for DAJE B. The G235M and G395M spectra are fitted jointly on H$\beta$ with the same intrinsic model, and only the model evaluated with the G395M LSF is plotted. The absorption component is statistically required ($\approx7.6\sigma$) on the $R\simeq1000$ spectrum, and the fitted properties are consistent with those from BlackTHUNDER, indicating that they are more likely real.}
\label{fig:absorbHb}
\end{figure}

\section{Comparison among different Balmer-line fitting settings}
\label{append:compare_fit}
\begin{figure*}
\centering
\includegraphics[width=\linewidth]{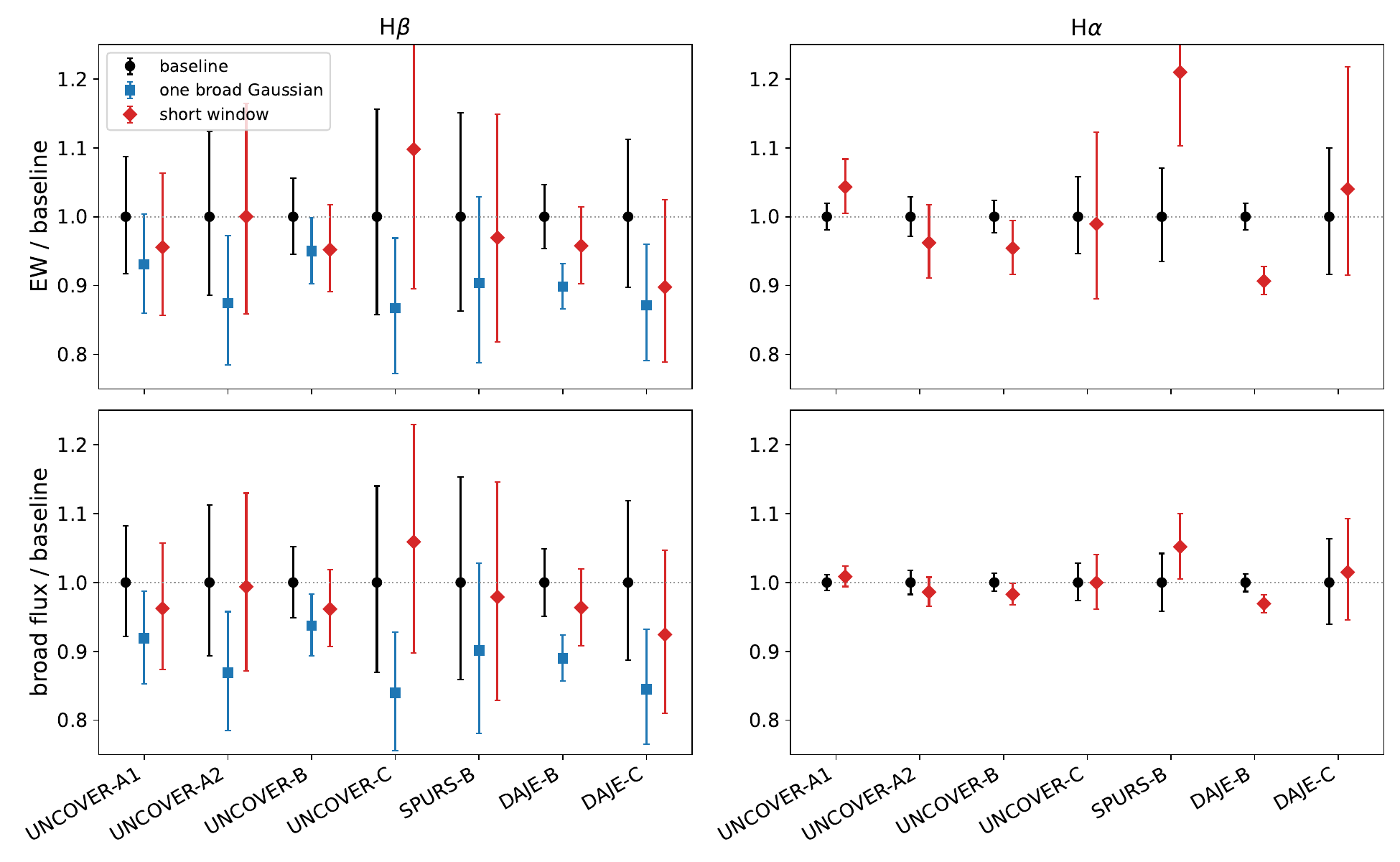}
\caption{Effect of the fitting setup on the measured EWs (top) and broad-line fluxes (bottom) of H$\beta$ (left) and H$\alpha$ (right). Each value is divided by the main-text results (Table~\ref{tab:lines}) of the same spectrum, so that the black circles lie at unity (dotted line) and their error bars show the statistical uncertainties of the main-text results. Blue squares are the fits with a single broad Gaussian, done for H$\beta$ only, and red diamonds are the fits within the short windows. Error bars are the 16th--84th percentiles of the posteriors.}
\label{fig:compare_fit}
\end{figure*}

In this appendix, we test how the fitting setup affects the measured H$\beta$ and H$\alpha$ properties in two ways. First, H$\beta$ can be fitted acceptably well by a single broad Gaussian; we adopt two in the main text for consistency with H$\alpha$, whose broad component requires two Gaussians (e.g., \citealt{furtak2025, deugenio2025}). We therefore refit H$\beta$ with one broad and one narrow Gaussian. Second, we narrow the fitting windows to $\pm10000~\mathrm{km~s^{-1}}$ around H$\alpha$ and to between $-10000$ and $+15000~\mathrm{km~s^{-1}}$ around H$\beta$, where the red limit is extended to include [\iona{O}{iii}] $\lambda5007$ and the continuum beyond it. These windows are about 30\% (H$\alpha$) and 50\% (H$\beta$) as wide as those of the main text. All other steps of the fits are unchanged. Figure~\ref{fig:compare_fit} compares the results with those of the main text.

As Figure~\ref{fig:compare_fit} shows, the H$\beta$ EW, the H$\beta$ flux, and the H$\alpha$ flux remain stable, mostly within $1\sigma$. Slight changes up to $\approx5\%-10\%$ are mainly decreases: a single Gaussian carries less flux in the wings than two, and the continuum fitted in the short windows is slightly higher under the lines, so part of the far wings may be attributed to it. The H$\alpha$ EW, however, changes beyond statistical uncertainties in the short-window fitting. Since the H$\alpha$ flux remains roughly stable, the EW difference arises mainly from the continuum level under the line. The fit statistics over the pixels in common, however, do not favor either window. The difference thus reflects systematic uncertainties of $\approx10\%$ for the H$\alpha$ EW and $\approx5\%$ for the other quantities. We adopt these differences as systematic uncertainties of the EWs in Section~\ref{sec:var}.

\section{Microlensing and millilensing calculations}
\label{append:lensing}

\subsection{Microlensing}
Image A of QSO1 is brighter in F150W than predicted by the lens model, by $\approx0.4$\,mag relative to B and C, while the B/C flux ratio agrees better with the model; we first test whether microlensing by intracluster stars can produce such an excess.

We take the convergence and shear at each observed image position from the lens model ($\kappa = 0.57$, $0.81$, $0.45$ and $\gamma = 0.14$, $0.44$, $0.06$ for A, B and C) and estimate the stellar convergence $\kappa_\star$ from the cluster light: the F150W \textsc{galfit} model of the cluster galaxies, averaged within $0\farcs1$ of each image, is converted to a stellar surface density using the solar absolute magnitude in the corresponding rest-frame band F115W \citep{Willmer2018} and a near-infrared mass-to-light ratio $M/L = 1$, as appropriate for an old stellar population, and divided by the critical surface density. 
This gives $\kappa_\star = 2.0\times10^{-4}$, $4.2\times10^{-3}$ and $3.9\times10^{-4}$, i.e. at most 0.5 \% of the total convergence. 
With stellar masses drawn from a \citet{Kroupa2001} mass function between $0.08$ and $1\,M_\odot$, the Einstein radius in the source plane is $R_E = 9.1$ light-days, and we compute magnification maps of $16000^2$ pixels containing $\gtrsim 200$ stars each with \textsc{SLSim} \citep{Khadka2026}, which uses the GPU code of \citet{Weisenbach2025}. 
Because microlensing statistics depend mainly on the half-light radius $R_{1/2}$ rather than the detailed source profile \citep{Mortonson2005}, we convolve the maps with Gaussian sources of $R_{1/2} = 0.5$--$30$ light-days for the continuum and $10$--$100$ light-days for the broad-line region (BLR); the narrow-line region is far too large to be affected ($\gtrsim 100$\,pc; \citealt{juodzbalis2026}). 

We found that images A and C, with so few stars, are essentially unaffected, with a scatter in $\Delta m$ below $0.03$\,mag for every source size, whereas B, the saddle-point image with the highest $\kappa_\star$, has a scatter of $\approx 0.2$\,mag for $R_{1/2} \lesssim 2$ light-days that falls below $0.1$\,mag for the BLR. 
An A$-$B offset as large as observed therefore occurs in about 4 \% of realizations for a compact continuum, almost entirely through demagnification of B, falling to about 1 \% at $R_{1/2} \approx 8$ light-days and below 0.5 \% for the BLR, while an A$-$C offset as large as observed occurs in at most 0.1 \% for any source size. 
The probability of a 0.3--0.4\,mag excess in A while B and C stay within 0.1\,mag of the model is $\lesssim 0.1$ \% for every source size, so microlensing by
intracluster stars cannot explain the excess brightness of image A.

\subsection{Millilensing}
We compute the expected millilensing signal in CDM by forward modeling the effects of cluster subhalo and line-of-sight halo populations. We render (sub)halo populations with {\tt{pyHalo}} \citep{Gilman++20}, with (sub)halos distributed within 1.5 arcsec of each lensed image. We model the line-of-sight halo mass function by drawing halo masses from the mass function of \citet{Sheth++01} and the concentration-mass relation of \citet{DiemerJoyce19}. We draw cluster subhalo masses from a mass function of the form 
\begin{equation}
\frac{d^2N}{dm dA} = \frac{\Sigma_{\rm{sub}}}{m_0} \left(\frac{m}{m_0}\right)^\alpha
\end{equation}
where $m$ is the subhalo mass at infall, $A$ the projected area in the lens plane, $m_0=10^8\,M_{\odot}$ is a pivot mass and $\alpha = -1.9$ the logarithmic slope. The parameter $\Sigma_{\rm{sub}}$, with units of number per kpc$^{2}$, sets the normalization of the mass function for subhalos at infall. We predict the evolved density profiles of subhalos using the tidal evolution model of \citet{Du++25}, modified for a cluster environment as in \citet{Perera++26}. The tidal evolution model predicts a typical bound mass fraction for cluster subhalos $f_{\rm{b}} \sim 0.1$, which gives the density profile of tidally-evolved subhalos according to the tidal tracks \citep{Du++24}. 

We simulate subhalo populations with $\log_{10} \left(\Sigma_{\rm{sub}} / \rm{kpc^{-2}}\right) = \left(-1.5, -1.0, -0.5, 0.0\right)$ to quantify how the millilensing signal depends on strength of perturbations by subhalos, and assumptions that enter the modeling of this population, including the abundance near lensed images around the cluster and the tidal evolution model. The normalization of the subhalo mass function sets the projected mass density in subhalos per unit area: 
\begin{equation}
\Sigma \equiv\int_{m_{\rm{L}}}^{m_{\rm H}} m\,\frac{d^2N}{dm\,dA}\,dm \approx 12.3\,m_0\,\Sigma_{\rm sub}, 
\end{equation}
with a minimum (maximum) halo mass at infall $m_{\rm L}=10^6 M_{\odot}$ ($m_{\rm H}=10^{10.7} M_{\odot}$). Assuming dark matter comprises a fraction $\left(1-f_{\rm{bar}}\right) \kappa$ of the total convergence, $\kappa \sim 0.6$, near lensed images, where $f_{\rm{bar}}$ is the projected baryon density, we can write the projected mass fraction in subhalos, $f_{\rm{sub}}$, as 
\begin{equation}
f_{\rm sub} = \frac{\Sigma_{\rm }}{\Sigma_{\rm DM}} = \frac{12.3\,f_{\rm b}\,m_0\,\Sigma_{\rm sub}}
{(1 - f_{\rm bar})\,\kappa\,\Sigma_{\rm crit}}
\approx \frac{0.059}{(1 - f_{\rm bar})\,\kappa}\;\frac{\Sigma_{\rm sub}}{{\rm kpc}^{-2}}.
\end{equation}
where $\Sigma_{\rm{crit}}$ is the critical density. The $\Sigma_{\rm{sub}}$ values we consider, for subhalos in the infall mass range $10^6 - 10^{10.7} M_{\odot}$, therefore correspond to $f_{\rm{sub}}$ values of $f_{\rm sub} \approx 0.4\%$, $1.2\%$, $3.9\%$, and $12\%$. 

For each $\Sigma_{\rm{sub}}$ we compute the distribution of magnifications for a point source, a finite-size source modeled as a circular Gaussian with a full width at half maximum of 20 and 100 pc. We find that the effect of the source size is negligible compared to the effect of changing $\Sigma_{\rm{sub}}$. 

Given the source redshift $z_{\rm{source}} = 7$, line-of-sight halos produce a level of perturbation comparable to or exceeding the subhalo population. With $\Sigma_{\rm{sub}}=0$, line-of-sight halos impart perturbations to the magnifications of $\sim 5$--$9\%$ (the standard deviation of $\ln \mu$, over the range spanned by the three images). For $\log_{10} \left(\Sigma_{\rm{sub}} / \rm{kpc^{-2}}\right) = \left(-1.5, -1.0, -0.5, 0.0\right)$, millilensing imparts perturbations of $6$--$9\%$, $6$--$10\%$, $8$--$14\%$, and $12$--$18\%$, respectively.

\section{Photoionization model grids}
\label{append:cloudy}

\begin{figure*}
\centering
\includegraphics[width=\textwidth]{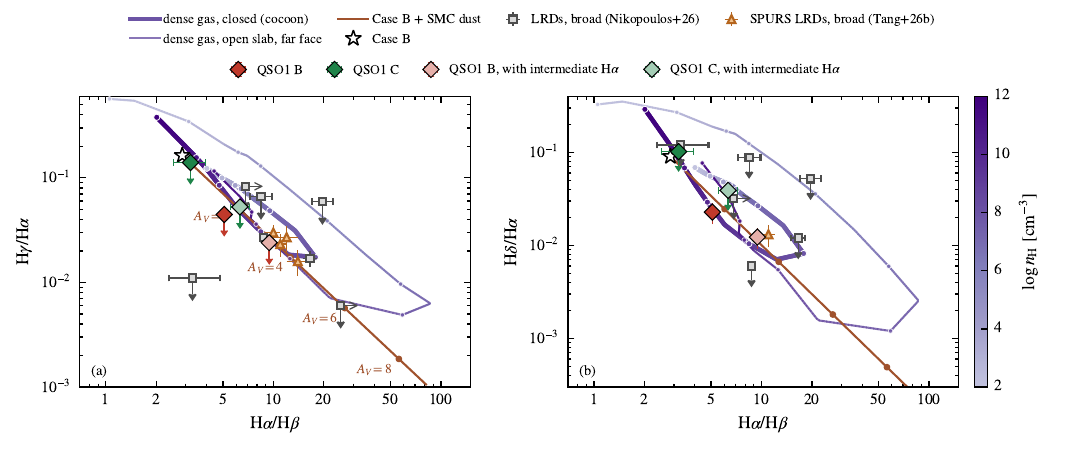}
\caption{Broad Balmer line ratios. H$\gamma$/H$\alpha$ (a) and H$\delta$/H$\alpha$ (b), against H$\alpha$/H$\beta$. The star represents Case~B, and the
associated brown track is Case~B reddened by a dust screen with the SMC extinction curve
\citep{gordon2003}, with dots every 2\,mag in $A_V$. The purple curves are our
dust-free \textsc{Cloudy} models for $N_{\rm H}=10^{24}$\,cm$^{-2}$, colored by
hydrogen density, with dots at integer $\log n_{\rm H}$: thick for a closed
geometry, thin for the light that leaves the shielded face of an open slab. 
Squares are the broad components of LRDs from \citet{nikopoulos2026b}, with
arrows for limits; triangles are the SPURS LRDs of \citet{tang2026b}; diamonds are the images of QSO1.}
\label{fig:balmer}
\end{figure*}

To convert the decrement into an extinction we need its intrinsic value. The
value usually adopted, 2.86, is that of Case~B recombination at
$T_{\rm e}=10^{4}$\,K and $n_{\rm e}=10^{2}$\,cm$^{-3}$
\citep{osterbrock2006}. It assumes that the Balmer lines are optically thin
and that collisional excitation is negligible, and both assumptions fail in
dense gas. If the excess over 2.86 is due to dust, the $A_V$ derived from
H$\alpha$/H$\beta$ predicts H$\gamma$/H$\alpha$ and H$\delta$/H$\alpha$
through the extinction curve. Dense gas follows that prediction only in some
regimes. To find them, we computed the line ratios of dust-free gas
illuminated by an AGN with \textsc{Cloudy} C23.01 \citep{gunasekera2023},
with the setup of \citet{yan2026}: ionization parameter
$\log U=-1.5$, metallicity $0.1\,Z_\odot$,
$n_{\rm H}=10^{2}$--$10^{12}$\,cm$^{-3}$ and column densities
$N_{\rm H}=10^{22}$--$10^{24}$\,cm$^{-2}$ (Figure~\ref{fig:balmer}). The decrement changes by less than 15\% between 0.01 and $1\,Z_\odot$
\citep{yan2026}, so the metallicity of the gas has little effect.

The ratios depend on the geometry. In a closed geometry the gas surrounds the
source, as in the cocoon models, and all the line photons leave through its
outer surface. In an open slab the lines leave through both faces, and the
light that leaves through the shielded face has crossed the whole column. The
light from the shielded face reproduces the decrements tabulated by
\citet{yan2026}, which reach $\simeq100$ at
$n_{\rm H}\simeq10^{7}$--$10^{8}$\,cm$^{-3}$. In the closed geometry the
decrement does not exceed $\simeq20$ on our grid, comparable to the limit of
$\simeq13$ for dust-free gas found by \citet{kiyota2026}.

The ratios also depend on the density. Below $n_{\rm H}\simeq10^{4}$\,cm$^{-3}$,
and for $N_{\rm H}\lesssim10^{23}$\,cm$^{-2}$, they are those of Case~B.
Between $10^{5}$ and $10^{8}$\,cm$^{-3}$ the Balmer lines are optically thick,
and at a given H$\alpha$/H$\beta$ the models give H$\gamma$/H$\alpha$ and
H$\delta$/H$\alpha$ above the values for dust, by factors of 1.3--2.8 in the
closed geometry and 3--14 from the shielded face. Above
$\simeq10^{8.5}$\,cm$^{-3}$ the level populations approach their LTE values,
and in both geometries the ratios lie within a factor of $\simeq1.5$ of the
reddening curve \citep{yan2026}. In this regime dust and dense gas give the same ratios. Close to LTE, above $\sim10^{10}$\,cm$^{-3}$,
the models are less certain, because the treatment of highly excited levels in
\textsc{Cloudy} is still being developed \citep{gunasekera2025}.

We computed a grid of \textsc{Cloudy} C23.01 models \citep{gunasekera2023}
of ionization-bounded, dust-free gas with the solar abundance pattern of
\citet{grevesse2010} scaled by $Z$, for $\log(Z/Z_\odot)=-3$ to 0 in steps of 0.25,
$\log U=-3.5$ to $-0.5$ in steps of 0.5 and
$n_{\rm H}=10^{2}$--$10^{6}$\,cm$^{-3}$ in steps of 1\,dex. We used
four ionizing spectra: the AGN continuum of Section~\ref{sec:decrement}; the
same with the X-rays reduced, $\alpha_{\rm ox}=-2$, as LRDs are X-ray weak;
and BPASS v2.2.1 binary populations of 3 and 10\,Myr \citep{eldridge2017,
stanway2018}, with the stellar metallicity equal to that of the gas. Around the solution we refined the grid to steps of 0.1 in $\log Z$ and
0.25 in $\log U$, for $\log(Z/Z_\odot)=-2.8$ to $-1.6$ and, at
$\log U\le-2.5$, up to $-1.0$, for the AGN and the 3\,Myr population at
$n_{\rm H}=10^{2}$ and $10^{4}$\,cm$^{-3}$. We fit
all the measured ratios together and quote the range of O/H over the models
within $\Delta\chi^2=2.3$ of the best fit.

Each ratio constrains a different parameter. At fixed $R_3$, O/H depends
mostly on $U$: $R_3=0.38$ gives $12+\log({\rm O/H})$ from 6.1 at $\log U=-0.5$ to 6.9 at $\log U=-3$ at $n_{\rm H}=10^{2}$\,cm$^{-3}$ (Figure~\ref{fig:metal}a). [O\,\textsc{ii}]$\lambda3727$/H$\beta$
sets $U$, since it falls by more than two orders of magnitude over this
range (Figure~\ref{fig:metal}b). It also selects between the two values of O/H that give the same
$R_3$, since the high-metallicity solution has strong [O\,\textsc{ii}]. The
ionizing spectrum changes O/H by up to 0.3\,dex at fixed $U$, and it sets
narrow He\,\textsc{ii}\,$\lambda4686$/H$\beta$: 0.09--0.14 for the AGN,
0.01--0.02 for the X-ray-weak AGN, and $\lesssim0.005$ for the stellar
populations. Density has little effect up to $10^{4}$\,cm$^{-3}$. At
$10^{6}$\,cm$^{-3}$ collisions suppress $\lambda5007$, and the same $R_3$
requires 0.3--0.7\,dex more oxygen \citep{hsiao2026}; such gas would have
[O\,\textsc{iii}]$\lambda4363$/H$\gamma\simeq0.1$--0.2, against 0.02--0.07 at
$10^{2}$--$10^{4}$\,cm$^{-3}$. At $R_3=0.55$ the calibration of
\citet{sanders2024} corresponds to $\log U\simeq-1.5$ to $-1$. Below
$12+\log({\rm O/H})=7.4$, the lower end of its calibration sample, the
calibration is extrapolated \citep{sanders2024, maiolino2025pristine}, and it
falls more steeply than any of the models (Figure~\ref{fig:metal}). Strong-line calibrations such as that of \citet{sanders2024} are
derived from star-forming galaxies, and assume gas at low density,
ionized by stars, with the ionization parameter $U$ of those galaxies;
these assumptions have not been tested for QSO1. The direct method, which
derives the electron temperature from
[O\,\textsc{iii}]$\lambda4363$/$\lambda5007$, is not feasible here owing
to the faintness of $\lambda4363$: in photoionization models with
$Z\simeq5\times10^{-3}\,Z_\odot$ it is 1--3\% of narrow H$\beta$, and it
lies on the red wing of broad H$\gamma$.

\bibliography{citations}{}
\bibliographystyle{aasjournalv7}

\end{document}